\documentclass[superscriptaddress,nofootinbib,a4paper,11pt]{article}
\usepackage{jheppub}
\usepackage{graphicx}
\usepackage{subcaption}
\usepackage{float}
\usepackage{amsmath}
\usepackage{amssymb}
\usepackage[utf8]{inputenc}
\usepackage{hyperref}
\usepackage{color}
\usepackage{slashed}
\usepackage{multirow}
\usepackage{enumitem}
\usepackage{lineno}
\allowdisplaybreaks

\newcommand{\be}{\begin{equation}}
\newcommand{\ee}{\end{equation}}
\newcommand{\thetaq}{\theta(\mathcal{Q} \bar{\mathcal{Q}})}

\begin{document}

\preprint{UCI-TR-2024-04}

\title{\boldmath Discovering Quirks through Timing at FASER and Future Forward Experiments at the LHC}

\author[b]{Jonathan L.~Feng,}
\author[a]{Jinmian Li,}
\author[a]{Xufei Liao,} 
\author[a]{Jian Ni,}
\author[c,d]{and Junle Pei}

\affiliation[a]{College of Physics, Sichuan University, Chengdu 610065, China}
\affiliation[b]{Department of Physics and Astronomy, University of California, Irvine, CA 92697 USA}
\affiliation[c]{Institute of High Energy Physics, Chinese Academy of Sciences, Beijing 100049, China}
\affiliation[d]{Spallation Neutron Source Science Center, Dongguan 523803, China}

\emailAdd{jlf@uci.edu}
\emailAdd{jmli@scu.edu.cn}
\emailAdd{peijunle@ihep.ac.cn}
\emailAdd{xufeiliao12@outlook.com}
\emailAdd{jian.ni.2020@outlook.com}

\abstract{
Quirks are generic predictions of strongly-coupled dark sectors.  For weak-scale masses and a broad range of confining scales in the dark sector, quirks can be discovered only at the energy frontier, but quirk--anti-quirk pairs are produced with unusual signatures at low $p_T$, making them difficult to detect at the large LHC detectors. We determine the prospects for discovering quirks using timing information at FASER, FASER2, and an ``ultimate detector'' in the far-forward region at the LHC.  NLO QCD corrections are incorporated in the simulation of quirk production, which can significantly increase the production rate. To accurately propagate quirk pairs from the ATLAS interaction point to the forward detectors, the ionization energy loss of charged quirks traveling through matter, the radiation of infracolor glueballs and QCD hadrons during quirk pair oscillations, and the annihilation of quirkonium are properly considered. The quirk signal is separated from the large muon background using timing information from scintillator detectors by requiring either two coincident delayed tracks, based on arrival times at the detector, or two coincident slow tracks, based on time differences between hits in the front and back scintillators.  We find that simple cuts preserve much of the signal, but reduce the muon background to negligible levels. With the data already collected, FASER can discover quirks in currently unconstrained parameter space.  FASER2, running at the Forward Physics Facility during the HL-LHC era, will greatly extend this reach, probing the TeV-scale quirk masses motivated by the gauge hierarchy problem for the broad range of dark-sector confining scales between 100~eV and 100~keV. 
}

\maketitle

%%%%%%%%%%%%%%%%%%%%%%%%%%%%%%%%%%%%%%%%
\section{Introduction} \label{sec:intro} 
%%%%%%%%%%%%%%%%%%%%%%%%%%%%%%%%%%%%%%%%

The Large Hadron Collider (LHC) at CERN is a powerful tool for scrutinizing the Standard Model (SM) of particle physics. In the years since the discovery of the Higgs boson~\cite{ATLAS:2012yve, CMS:2012qbp}, no evidence for beyond-the-SM (BSM) physics has been found. It is therefore essential to carefully consider the possibility that BSM particles have distinct characteristics that have caused them to be inadvertently overlooked during the conventional event reconstruction process.  In fact, such particles are motivated by everything from dark sectors~\cite{Essig:2013lka,Gori:2022vri} to neutral naturalness~\cite{Chacko:2005pe, Burdman:2006tz, Burdman:2008ek,Cai:2008au, Curtin:2015bka, Craig:2015pha, Serra:2019omd, Batell:2022pzc}, and their generic presence in BSM models has inspired searches for a wide variety of new signatures, including long-lived particles~\cite{Lee:2018pag,Beacham:2019nyx,Alimena:2019zri,Anchordoqui:2021ghd,Feng:2022inv,Knapen:2022afb,Garzelli:2022vad}, disappearing tracks~\cite{Chen:1996ap, Feng:1999fu, CMS:2018rea, ATLAS:2022rme, CMS:2020atg, CMS:2023mny}, emerging jets~\cite{Schwaller:2015gea, CMS:2018bvr}, soft bombs~\cite{Knapen:2016hky}, and many others. 

In this study, we consider quirks $\mathcal{Q}$~\cite{Kang:2008ea}, which can have fascinating signatures.  Quirks are present in models where the SM is extended to a dark sector. If the dark sector includes an Abelian gauge symmetry, the predictions for observable phenomena include dark photons and milli-charged particles~\cite{Holdom:1985ag}, both of which have been studied extensively.  Much less well studied, but perhaps even more generic, is that the dark sector has a non-Abelian gauge symmetry.  We will refer to this dark force as infracolor and denote its confinement scale as $\Lambda$, analogous to $\Lambda_{\text{QCD}}$, the confinement scale of QCD in the SM.  

Quirks are matter particles that have infracolor and are also charged under a SM gauge group.  Given their SM charge, quirks may be pair-produced in large numbers at colliders.  At the same time, given their infracolor, quirks are connected by an infracolor string, and, if $\Lambda$ is less than the mass of the lightest quirk, the process of breaking the infracolor string by producing an additional quirk pair out of the vacuum is highly suppressed.  Instead, a macroscopic gauge flux tube is formed, connecting the $\mathcal{Q} \bar{\mathcal{Q}}$ pair, and there is an attractive infracolor force between the quirk and anti-quirk proportional to $\Lambda^2$. The quirk and anti-quirk then appear as minimum-ionizing particles (MIPs), but they oscillate around their center-of-mass (CoM), producing tracks that can be spectacularly different from those of SM particles~\cite{Li:2019wce, Li:2020aoq}.  

In addition to their presence in generic strongly-coupled dark sectors, quirks are further motivated by models of neutral naturalness~\cite{Chacko:2005pe, Burdman:2006tz, Burdman:2008ek,Cai:2008au, Curtin:2015bka, Craig:2015pha, Serra:2019omd, Batell:2022pzc}.  In these models, the gauge hierarchy problem is ameliorated by the existence of non-colored top partners charged under an additional gauge group.  These models predict quirks with weak-scale masses, beyond current bounds, but potentially within the reach of LHC searches.  Given the increasingly stringent bounds on colored top partners, there is increasing attention focused on neutral naturalness, adding motivation for searches for quirks with masses between 100 GeV and 1 TeV over the full range of possible $\Lambda$.  

In this study, we determine the prospects for discovering quirks at forward detectors at the LHC.  Because $\mathcal{Q}\bar{\mathcal{Q}}$ pairs have very little transverse momentum $p_T$, once produced at the LHC, they preferentially travel in the forward direction, making forward detectors a natural place to search for them.  We consider two detectors:
FASER~\cite{FASER:2018ceo, FASER:2018bac, FASER:2022hcn, FASER:2018eoc}, a detector located along the beam collision axis or line of sight (LOS), 480 m east of the ATLAS interaction point (IP), which has been collecting data at LHC Run 3 since 2022, and  FASER2~\cite{Feng:2022inv}, an experiment proposed to operate in the Forward Physics Facility (FPF)~\cite{Anchordoqui:2021ghd, Feng:2022inv} during the High-Luminosity LHC (HL-LHC) era.  In addition, we consider an ``ultimate detector'' (UD), consisting of scintillators that cover the front and back walls of the FPF.  The UD is an idea invented for this study and is not at all well defined experimentally, but it is meant to represent the theoretically best possible sensitivity achievable by a detector at the FPF.  

The possibility of detecting quirks at FASER and FASER2 was previously considered by two of us and collaborators in Refs.~\cite{Li:2021tsy, Li:2023jrt}.  In those studies, the expected numbers of quirk events that can reach the FASER and FASER2 detectors were calculated by carefully simulating the motion of quirks as they pass through the LHC infrastructure between the ATLAS IP and the detectors, providing insights into the possible signal rates at FASER and FASER2.  In this study, we address the problem of differentiating the quirk signal from the large background of high-energy muons and determine the prospects for discovery.  

To do this, we exploit the fact that quirks are heavy and so are often produced at velocities considerably below the speed of light.  We therefore consider two sets of criteria using timing to separate the quirk signal from the muon background:
\begin{itemize}
[itemsep=0.03cm, topsep=0.15cm, leftmargin=1.3em]
\item {\em Delayed Tracks}.  With information about the bunch crossing times at the ATLAS IP, one can predict when muons from the ATLAS IP will pass through FASER.  Quirks can be selected by looking for two coincident tracks in FASER that are significantly delayed relative to this time.  This delayed track selection makes use of a feature of far-forward detectors that has not been exploited previously for BSM searches, namely, their long distance from the ATLAS IP, which amplifies the time delay relative to conventional detectors. 
\item {\em Slow Tracks}. Without coordination with the LHC clock, the time difference between hits in the front and back scintillators can be measured. Quirks can be separated from muons by requiring two coincident slow tracks, where, for each track, this time difference is significantly larger than typical of particles traveling at the speed of light.
\end{itemize}

As we will see, the sensitivity achieved through these two analyses is striking.  In both cases, experimentally reasonable criteria preserve most of the signal while reducing the muon background to a negligible level.  With the data currently available from the integrated luminosity of approximately $60~\text{fb}^{-1}$ collected in 2022 and 2023, FASER is already able to probe quirk parameter space beyond current bounds from other LHC experiments.  This sensitivity will be significantly extended with the full Run 3 luminosity.  At FASER2, the sensitivity extends well into the TeV mass range motivated by the gauge hierarchy problem for a range of $\Lambda$ spanning three orders of magnitude from 100~eV to 100~keV.  Finally, the FASER2 sensitivity is improved even further by the UD, motivating more serious consideration of a detector dedicated to the search for delayed or slow particles. 

This paper is structured as follows. In Sec.~\ref{sec:2}, we introduce four simple quirk models and give an overview of quirk parameter space, the possible LHC signals, and existing bounds.  In Sec.~\ref{sec:3}, we discuss the cross sections and kinematics of quirk production at the LHC. Sec.~\ref{sec:4} provides the details for determining the lifetime of the quirk pair after production, including the effects of radiative energy loss and quirkonium annihilation. Sec.~\ref{sec:5} focuses on determining the trajectories of quirks as they travel from the ATLAS IP to the forward detectors. Finally, the sensitivity reaches of all detectors for the four quirk models are presented in Sec.~\ref{sec:6}, and we collect our conclusions in Sec.~\ref{sec:7}.

%%%%%%%%%%%%%%%%%%%%%%%%%%%%%%%%%%%%%%%%
\section{Quirk Models and Existing Bounds} \label{sec:2}
%%%%%%%%%%%%%%%%%%%%%%%%%%%%%%%%%%%%%%%%

In this work, we define four simple quirk models to study the discovery prospects of FASER, FASER2, and the UD. Each simplified model contains a fermionic/scalar quirk that is in the fundamental representation of the infracolor $\text{SU}(N_{\text{IC}})$ group, but has the same SM quantum numbers as either the right-handed charged leptons or the right-handed down-type quarks.  Under the $\text{SU}(N_{\text{IC}}) \times \text{SU}_C(3) \times \text{SU}_L(2) \times \text{U}_Y(1)$ gauge symmetry, they are 
\begin{align}
	\mathcal{E} &= \left( N_{\text{IC}}, 1, 1, -1 \right) \label{eq::qn4}, \\
	\mathcal{D} &= \left( N_{\text{IC}}, 3, 1, -1/3 \right), \label{eq::qn3}\\
	\tilde{\mathcal{E}} &= \left( N_{\text{IC}}, 1, 1, -1 \right), \label{eq::qn2} \\
	\tilde{\mathcal{D}} &= \left( N_{\text{IC}}, 3, 1, -1/3 \right), \label{eq::qn1}
\end{align}
where $\mathcal{E}$ and $\mathcal{D}$ are fermionic quirks, and $\tilde{\mathcal{E}}$ and $\tilde{\mathcal{D}}$ are scalar quirks. The quirk production cross sections are proportional to $N_{\text{IC}}$, and the radiation of infracolor glueballs and the quirk annihilation rate, which will be discussed in detail in Sec.~\ref{sec:4}, also depend on $N_{\text{IC}}$. We will take $N_{\text{IC}}=2$ in this work. This choice is conservative with respect to the production cross section and allows us to compare to previous results where similar $N_{\text{IC}}$ and quirk quantum numbers were considered~\cite{Knapen:2017kly, Farina:2017cts, Evans:2018jmd}.

With this choice of the infracolor gauge group and quirk representations, each model is completely defined by two continuous parameters,
\begin{equation}
m_{\mathcal{Q}} \text{ and } \Lambda \ ,
\end{equation}
where $m_{\mathcal{Q}}$ is the quirk mass, and $\Lambda$ is the infracolor confinement scale, which is interchangeable with the value of the infracolor gauge coupling at some high energy scale.

The amplitude of quirk oscillations can be roughly estimated to be
\begin{align}
\ell \sim \frac{m_{\mathcal{Q}}}{\Lambda^2} \sim 1~\text{cm} \ \bigg[ \frac{1~\text{keV}}{\Lambda} \bigg]^2 \ \bigg[\frac{m_{\mathcal{Q}}}{100~\text{GeV}} \bigg] \ .
\label{eq:amplitude}
\end{align}
For the reasons noted in Sec.~\ref{sec:intro}, we will consider $m_{\mathcal{Q}}$ in the range between 100 GeV and a few TeV.  The quirk signatures at the LHC, along with the existing bounds, are, then, largely determined by the value of $\Lambda$, which can vary over many orders of magnitude:
\begin{itemize}
[itemsep=0.03cm, topsep=0.15cm, leftmargin=1.3em]

\item $\Lambda \lesssim 100~\text{eV}$, $\ell \gtrsim 1~\text{m}$. For extremely low values of $\Lambda$, the infracolor force is very weak, and the deviation of a charged quirk track from  a typical charged track is below what can typically be measured, given the finite spatial resolution of detectors. Since the typical track reconstruction allows a $\chi^2/$DOF value as large as 5~\cite{CMS-PAS-EXO-16-036}, the quirk signal with normally reconstructed tracks can be constrained using conventional searches for heavy stable charged particles (HSCP) at the LHC~\cite{CMS-PAS-EXO-16-036, ATLAS:2016onr, Farina:2017cts}.

\item $100~\text{eV} \lesssim \Lambda \lesssim 10~\text{keV}$, $1~\text{m} \gtrsim \ell \gtrsim 0.1~\text{mm}$. 
In this range of $\Lambda$, the infracolor force is stronger and has a pronounced effect on quirk trajectories to such an extent that the non-helical track of each quirk from the pair is typically disregarded during conventional event reconstruction at the LHC. However, if the quirk pair is produced alongside an energetic jet from initial state radiation (ISR), the event can be seen as missing transverse energy, as the quirks traverse the calorimeters without significant energy loss. This signal can be constrained through monojet searches conducted at the LHC~\cite{Farina:2017cts,CMS-PAS-EXO-16-037,ATLAS:2016bek}. Alternatively, if the recoiling jet is low in energy, the quirk-pair system, consisting of two charged quirks, may experience substantial ionization energy loss and eventually come to a halt. During periods with no active $pp$ collisions, the two quirks can annihilate. This particular signal is constrained by searches for stopped long-lived particles, or out-of-time (OoT) searches, at the LHC~\cite{Evans:2018jmd,ATLAS:2013whh,CMS:2017kku}.   Given that the infracolor force (for $\Lambda\sim \text{keV}$) is significantly stronger than the Lorentz force (with magnetic field strength $B\sim 1~\text{T}$), all the hits from the two charged quirks in a pair will be approximately located in a plane. In Refs.~\cite{Knapen:2017kly, Li:2019wce}, specific search methods were developed using this coplanarity and the relatively high ionization energy of quirk hits.

\item $10~\text{keV} \lesssim \Lambda \lesssim 10~\text{MeV}$, $0.1~\text{mm} \gtrsim \ell \gtrsim 0.1~\text{nm}$.  In this range, the quirk-pair system follows an almost straight track in the detector and is reconstructed as a single charged particle with high ionization energy loss ($dE/dx$). This signal was previously investigated at the Tevatron~\cite{D0:2010kkd}.

\item $\Lambda \gtrsim 10~\text{MeV}$, $0.1~\text{nm} \gtrsim \ell$. Finally, in this range, the infracolor force becomes strong, and the rapid oscillations of the quirk pair result in a high rate of radiation of hidden glueballs and SM particles. The two quirks then lose energy rapidly, fall into their ground state, and subsequently annihilate into dark sector or SM particles. In the latter case, these quirk signals can be detected by searching for resonances in the final states of SM particles~\cite{Cheung:2008ke, Harnik:2008ax, Harnik:2011mv, Fok:2011yc, Chacko:2015fbc, Capdevilla:2019zbx}.

\end{itemize}

From this brief tour of the quirk parameter space, we see that for $\Lambda$ below roughly 100 eV, quirks are so loosely bound that they do not preferentially travel in the forward direction and can be discovered with conventional LHC searches. As we will see, above the MeV scale, quirks oscillate rapidly, lose energy, and annihilate before reaching the forward detectors.  Given these basic constraints, then, without even considering signal and background rates, we see that the range of confinement scales that can be targeted at forward detectors is very roughly the four orders of magnitude between 100~eV and an MeV.

%%%%%%%%%%%%%%%%%%%%%%%%%%%%%%%%%%%%%%%%
\section{Quirk Production and Kinematics at the LHC} \label{sec:3}
%%%%%%%%%%%%%%%%%%%%%%%%%%%%%%%%%%%%%%%%

To simulate quirk production at the LHC, the \texttt{FeynRules}~\cite{Alloul:2013bka} package is used to write the four simple quirk models in Universal FeynRules Output format. 
Those models are imported into the \texttt{MG5\_aMC@NLO}~\cite{Alwall:2014hca} framework to conduct a Monte Carlo simulation of quirk events at the LHC, in which \texttt{Pythia8} is used for simulating the parton shower of both SM particles and colored quirks. 

The colored quirks $\mathcal{D}$ and $\tilde{\mathcal{D}}$ will hadronize.  This hadronization will not significantly change the momenta of the quirks, since the quirks are much heavier than the quarks participating in the hadronization, and so hadronization is turned off in our simulation. However, confinement with respect to SM QCD implies that only quirk-quark bound states with integer electric charges are observable at the detector.  In this study, we are only concerned with final states with non-zero electric charge.  A study using the \texttt{Pythia8}~\cite{Sjostrand:2007gs} simulation has found that the probability for the quirk-quark bound state to have charge $\pm 1$ is around 30\%~\cite{Knapen:2017kly}. To take this into account, we multiply the production cross sections for colored quirks by a factor of 1/3, and we will simply refer to the charge $\pm 1$ quirk-quark bound states as $\mathcal{D}$ or $\tilde{\mathcal{D}}$. The equations of motion (EoM) for the quirk-quark bound states are similar to the EoM for quirks, since the masses of the quirks are much larger than those of the quarks. 

\begin{figure}[tbp]
\begin{center}
\includegraphics[width=0.60\textwidth]{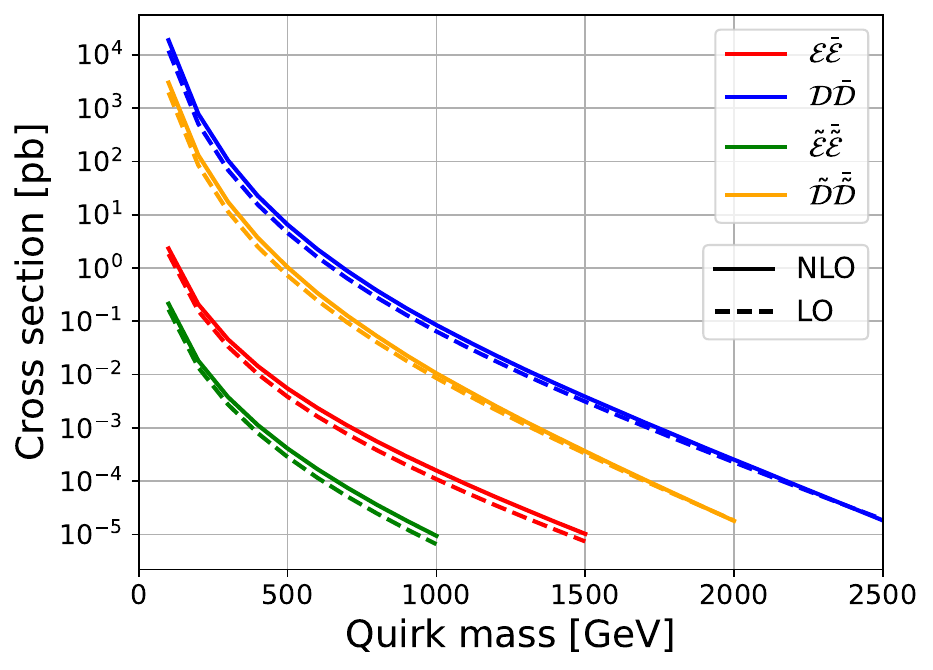}
\end{center}
 \vspace*{-0.15in}
\caption{The LO and NLO cross sections for quirk pair production at the 13 TeV LHC for the four quirk models considered. 
\label{fig::xsecs}}
\end{figure}

Because quirks have SM charges, they can be pair produced at the LHC. The color-neutral quirks $\mathcal{E}$ and $\tilde{\mathcal{E}}$ are produced mainly through the Drell-Yan process, and the colored quirks $\mathcal{D}$ and $\tilde{\mathcal{D}}$ are produced dominantly by the QCD process with an $s$-channel gluon.  In Figure~\ref{fig::xsecs} we show the production cross section of quirk pairs from $pp$ collisions at center-of-mass energy $\sqrt{s} = 13~\text{TeV}$, which we use throughout this work. The cross sections for $\sqrt{s} = 13.6~\text{TeV}$ at LHC Run 3 and the planned $\sqrt{s} = 14~\text{TeV}$ during the HL-LHC era are slightly larger and can be expected to improve our results slightly. Both the leading order (LO) and next-to-leading order (NLO) cross sections are calculated by \texttt{MG5\_aMC@NLO}~\cite{Alwall:2014hca}.  The NLO QCD corrections may increase the production rate by significant $K$-factors, with the enhancement most pronounced for heavier color-neutral quirks and lighter colored quirks.  For $m_{\mathcal{Q}} = 200~\text{GeV}$ (1~TeV), the $K$-factors for $\mathcal{E}$, $\mathcal{D}$, $\tilde{\mathcal{E}}$, and $\tilde{\mathcal{D}}$ are 1.33 (1.45), 1.54 (1.32), 1.34 (1.43), and 1.55 (1.23), respectively.  

\begin{figure}[tbp]
\begin{center}
\includegraphics[width=0.49\textwidth]{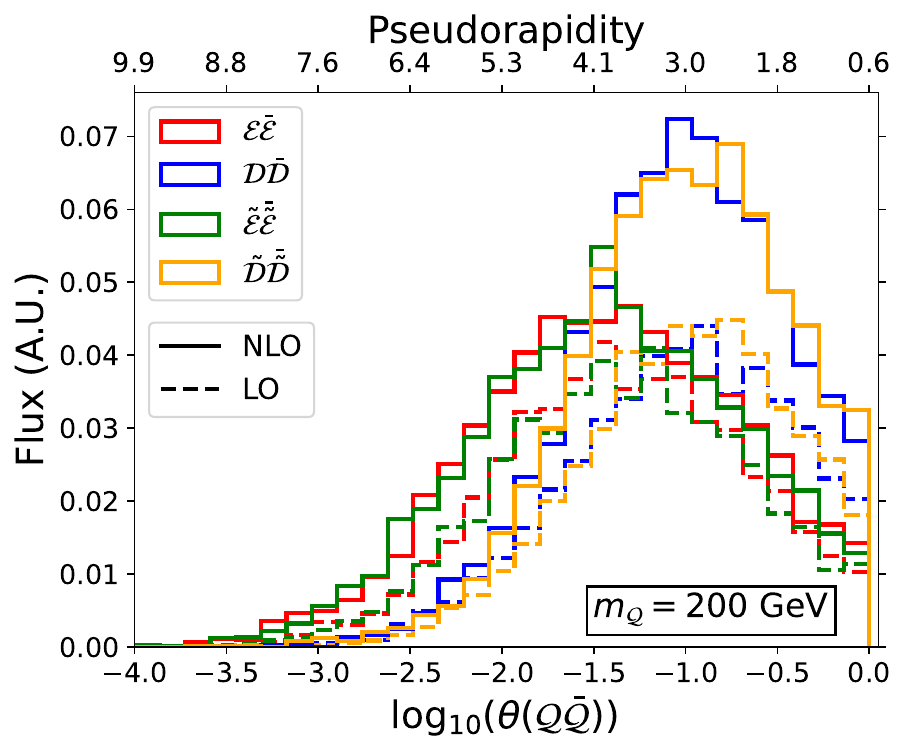} \hfill
\includegraphics[width=0.49\textwidth]{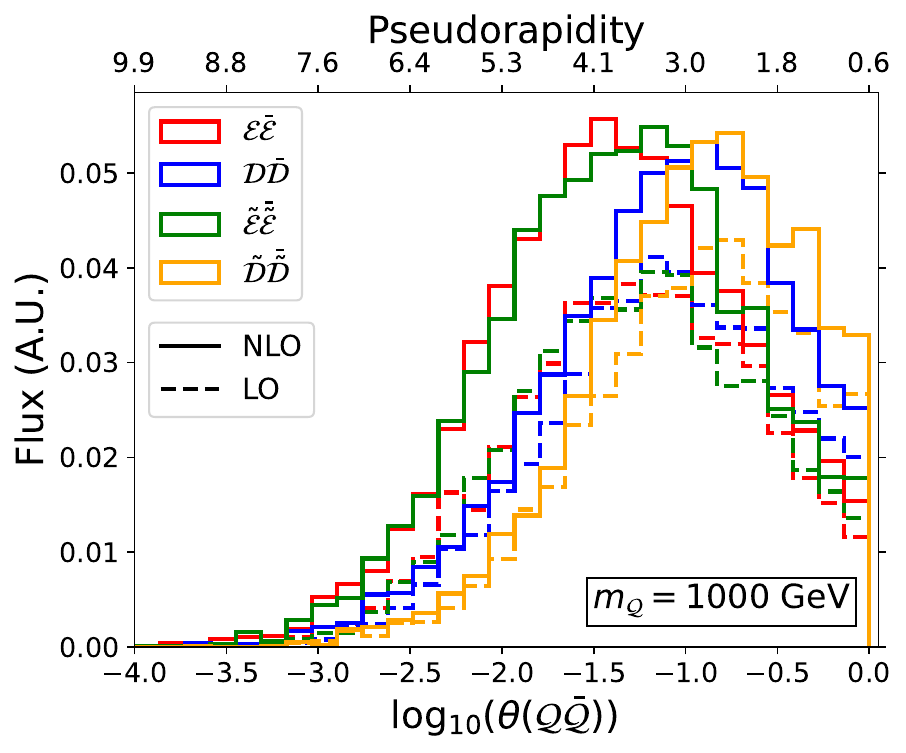}
\end{center}
 \vspace*{-0.15in}
\caption{Distributions of $\thetaq$, the polar angle relative to the beamline of the momentum of the quirk-pair system when produced.  Both LO and NLO results are shown for quirk masses of 200 GeV (left) and 1 TeV (right). 
\label{fig::polarangles}}
\end{figure}

For this study, the polar angle relative to the beamline of the momentum of the quirk-pair system when produced, which we denote $\thetaq$, is an important kinematic variable, as it controls the signal efficiency at the forward detectors. Non-zero values of $\thetaq$ result from ISR and final state radiation (FSR) in the quirk production processes.  In Figure~\ref{fig::polarangles}, we present the distributions of $\thetaq$ for the four quirk models as functions of quirk mass. The LO distributions are normalized to unity, while the NLO distributions are normalized to their corresponding $K$-factors. The fraction of forward (small $\thetaq$) events is higher for color-neutral quirks than for colored quirks, because the colored ones undergo more intense FSR, which deflects the quirk pair away from the beamline. For colored quirks with the same quantum numbers but different spin, the difference in the $\thetaq$ distributions is more significant for heavier quirks.

\begin{figure}[tbp]
	\begin{center}
		\includegraphics[width=0.49\textwidth]{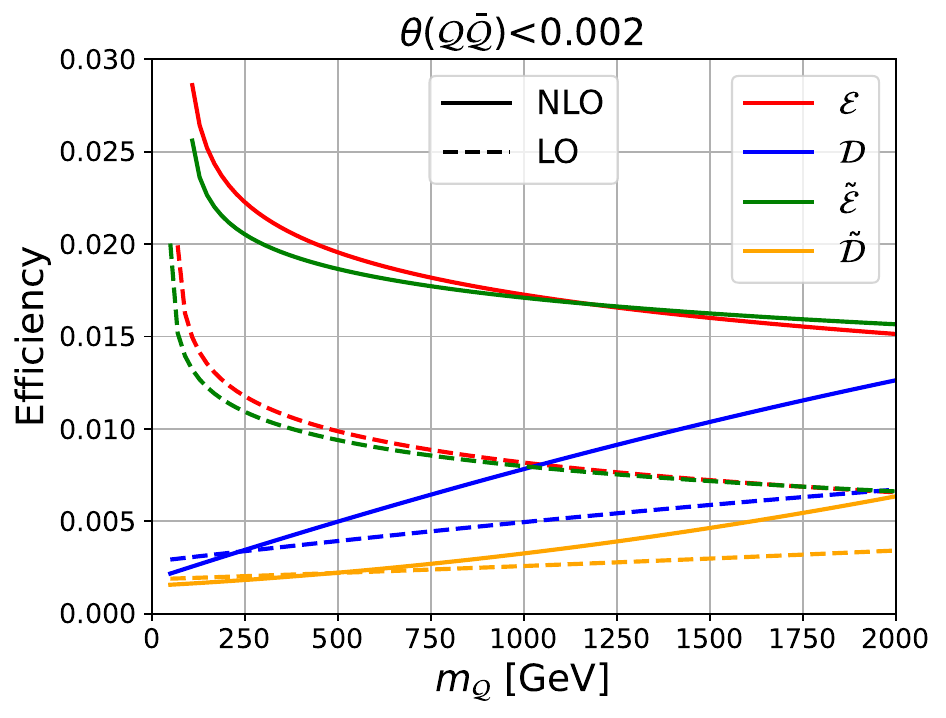}
  \hfill
		\includegraphics[width=0.49\textwidth]{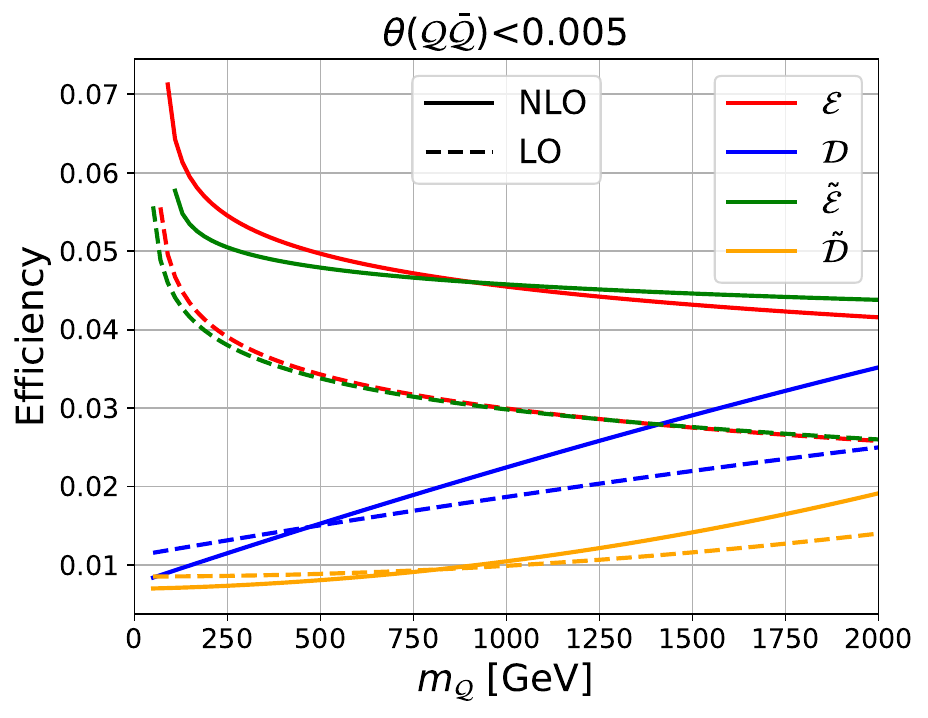}
	\end{center}
 \vspace*{-0.15in}
\caption{The fraction of events that have $\thetaq < 0.002$ (left) and 0.005 (right), where $\thetaq$ is the polar angle relative to the beamline of the quirk-pair system when produced.  Results are shown as functions of quirk mass $m_{\mathcal{Q}}$ for all four quirk models at both LO and NLO.}
\label{fig::effskins}
\end{figure}

The implications of these $\thetaq$ distributions for signal acceptance can be seen in Figure~\ref{fig::effskins}, where we plot the fractions of events that have $\thetaq < 0.002$ and $\thetaq < 0.005$ for the four quirks models as a function of quirk mass.  The requirement $\thetaq < 0.002$ is roughly the requirement that the quirk system when produced has a momentum that points into the FASER2 detector. We see that the fraction of quirk events that satisfy this requirement varies significantly for the different quirk types and masses, but is of the order of 1\%. The preference for forward production is pronounced: FASER2 covers only $6 \times 10^{-7}$ of the total solid angle, but $\sim 10^{-2}$ of quirk pairs are produced traveling in this direction.

The requirement $\thetaq < 0.005$ is dictated by the fact that for $\thetaq > 0.005$, quirk events are very unlikely to pass through FASER2.  However, for $\thetaq < 0.005$, quirks can still possibly pass through the FASER2 detector, even if the quirk system's initial momentum does not point into the FASER volume, given a macroscopic oscillation amplitude.  We see that the fraction of events with $\thetaq < 0.005$ is a factor of 2 to 3 times higher than for $\thetaq < 0.002$.

For colored quirks, the fraction of events that are within the detector acceptance is higher for larger masses, because FSR dominates over ISR, and FSR deflects the momentum direction less for heavier quirks.  On the other hand, for color-neutral quirks, only ISR is relevant, and the angular cut efficiencies are higher for lighter color-neutral quirks. Last, we see that for almost all cases, NLO QCD effects increase the angular cut efficiencies. Refs.~\cite{Cullen:2012eh, Backovic:2015soa, Ruiz:2015zca, Fuks:2016ftf} have reached similar conclusions for other BSM particles with the same quantum numbers. 

%%%%%%%%%%%%%%%%%%%%%%%%%%%%%%%%%%%%%%%%
\section{Radiative Energy Loss and Annihilation} \label{sec:4}
%%%%%%%%%%%%%%%%%%%%%%%%%%%%%%%%%%%%%%%%

Quirks are pair produced at the LHC with kinetic energy in the CoM frame of the quirk pair. This kinetic energy can be reduced by the radiation of infracolor glueballs, QCD hadrons, and SM photons as the quirks oscillate about their CoM. Through this radiation, the quirk pair may eventually settle into the quirkonium ground state, which can then lead to constituent quirk annihilation. The rates of radiative energy loss and quirk annihilation have been studied previously in Refs.~\cite{Evans:2018jmd, Li:2023jrt}.  Here we summarize these results and show that, for our study, the effect of radiative energy loss may become important for $\Lambda \gtrsim 100~\text{keV}$, and, if the quirks decay to their quirkonium ground state, they will immediately annihilate.  

As discussed in Ref.~\cite{Evans:2018jmd}, the radiation of infracolor glueballs is much more intense than SM photon radiation.  We therefore begin by considering the case of color-neutral quirks, where there is no QCD hadron radiation, and so energy loss through infracolor glueball radiation is clearly dominant. Assuming that during each period of quirk oscillation the quirk pair has a probability $\epsilon$ to emit an infracolor glueball with energy $\Lambda$~\cite{Kang:2008ea,Evans:2018jmd}, we find 
\begin{align}
\frac{dE}{dt} = - \frac{\epsilon \Lambda}{T}~,
\end{align}
where $E$ is the kinetic energy of the quirk in the CoM frame, and $T$ is the quirk's oscillation period.  Just after production, the quirk's oscillation amplitude is $r \gg \Lambda^{-1}$, given its large kinetic energy. For $\Lambda \gtrsim 1~\text{keV}$ ($\sim \text{\AA}^{-1}$), the interaction between two quirks is described by the linear potential $V(r) \sim \Lambda^2 r$, which gives the oscillation period $T=2\Lambda^{-2} \sqrt{2m_\mathcal{Q} E_0}$, where $E_0$ is the initial kinetic energy of the quirk in the CoM frame.  We can then determine the time required to reduce the kinetic energy from $E_0$ to $\sqrt{m_{\mathcal{Q}} \Lambda}$ (the kinetic energy corresponding to an oscillation amplitude around $\Lambda^{-1}$) to be 
\begin{align}
\tau^{\text{IC}}_{\text{linear}} \sim  \int_{E_0}^{\sqrt{m_\mathcal{Q} \Lambda}}\frac{dE}{dE/dt}=\frac{4 \sqrt{2 m_\mathcal{Q}} \left(E_0^{3 / 2}-\sqrt{m_\mathcal{Q} \Lambda}^{3/2}\right)}{3 \epsilon \Lambda^{3}}~.
\end{align}

The quirk potential becomes Coulombic, with potential $V(r) \sim -\alpha_{\text{IC}} (r)/r$, when the oscillation amplitude drops below $\Lambda^{-1}$, that is, when the kinetic energy drops below $\sqrt{m_{\mathcal{Q}} \Lambda}$.  Here the running coupling is $\alpha_{\text{IC}}(r) = 1/ [b_0 \log (1/(r^2 \Lambda^2))]$, where $b_0=\frac{1}{4\pi}(\frac{11}{3} N_{\text{IC}}- \frac{2}{3} N_f- \frac{1}{6} N_s)$. The numbers of fermions and scalars are $(N_f, N_s)=(1,0)$ for the $\mathcal{D}$, $\mathcal{E}$ scenarios and $(N_f, N_s)=(0,1)$ for the $\tilde{\mathcal{D}}$, $\tilde{\mathcal{E}}$ scenarios. In this case, the quirk oscillation period is given by $T=\pi\sqrt{m_\mathcal{Q} r^3 \alpha_{\text{IC}}^{-1}}$.  The rate of change of the binding energy $B=-V(r)$ is~\cite{Evans:2018jmd}
\begin{align}
\frac{dB}{dt}=\frac{\epsilon\Lambda B^{3/2}}{\pi \alpha_{\text{IC}} \sqrt{m_\mathcal{Q}}}~.
\end{align}
Given this, we find that the time it takes for the binding energy to change from $B_{\text{init}} \sim \Lambda$ to $B_{\text{final}} \sim \alpha_{\text{IC}}^2 m_{\mathcal{Q}}$ is 
\begin{align}
\tau^{\text{IC}}_{\text{Coulomb}}\sim \int_{B_{\text{init}}}^{B_{\text{final}}} \frac{dB}{dB/dt} =\frac{2\pi \alpha_{\text{IC}} \sqrt{m_\mathcal{Q}}}{\epsilon \Lambda^{3/2}} - \frac{2 \pi}{\epsilon \Lambda}~.~
\end{align}
Throughout parameter space,  $\tau^{\text{IC}}_{\text{linear}}$ is always larger than $\tau^{\text{IC}}_{\text{Coulomb}}$; the quirk-pair system spends most of its time in the linear regime. 

For colored quirks $\mathcal{D}$ and $\tilde{\mathcal{D}}$, the total time to evolve to the ground state through QCD hadron radiation can be determined in a similar way. Assuming that in each quirk oscillation the probability is $\epsilon^\prime$ to emit a QCD hadron with energy $E_{\text{QCD}} \sim 100$ MeV, we have 
\begin{align}
	\frac{dE}{dt} = -\frac{E_\text{QCD} \epsilon^\prime}{T}~. \label{eq:eloss2}
\end{align}
As a result, one can obtain
\begin{align}
	\tau^{\text{QCD}}_{\text{linear}} \sim  \int_{E_0}^{\sqrt{m_\mathcal{Q} \Lambda}}\frac{dE}{dE/dt}=\frac{4 \sqrt{2 m_\mathcal{Q}} \left(E_0^{3 / 2}-\sqrt{m_\mathcal{Q} \Lambda}^{3/2} \right)}{3 \epsilon^\prime E_\text{QCD} \Lambda^{2}} \label{eq:tlic2}
\end{align}
for the linear potential phase, and
\begin{align}
	\tau^{\text{QCD}}_{\text{Coulomb}} \sim \int_{B_{\text{init}}}^{B_{\text{final}}} \frac{dB}{dB/dt} = \frac{2\pi \alpha_{\text{IC}} \sqrt{m_\mathcal{Q}}}{\epsilon^\prime E_\text{QCD} \Lambda^{1/2}} - \frac{2 \pi}{ \epsilon^\prime E_\text{QCD}}
\end{align}
for the Coulombic potential phase.  As in the case of color-neutral quirks, $\tau^{\text{IC}}_{\text{linear}}$ is always larger than $\tau^{\text{IC}}_{\text{Coulomb}}$ for colored quirks.   

\begin{figure}[tbp]
\begin{center}
\includegraphics[width=0.49\textwidth]{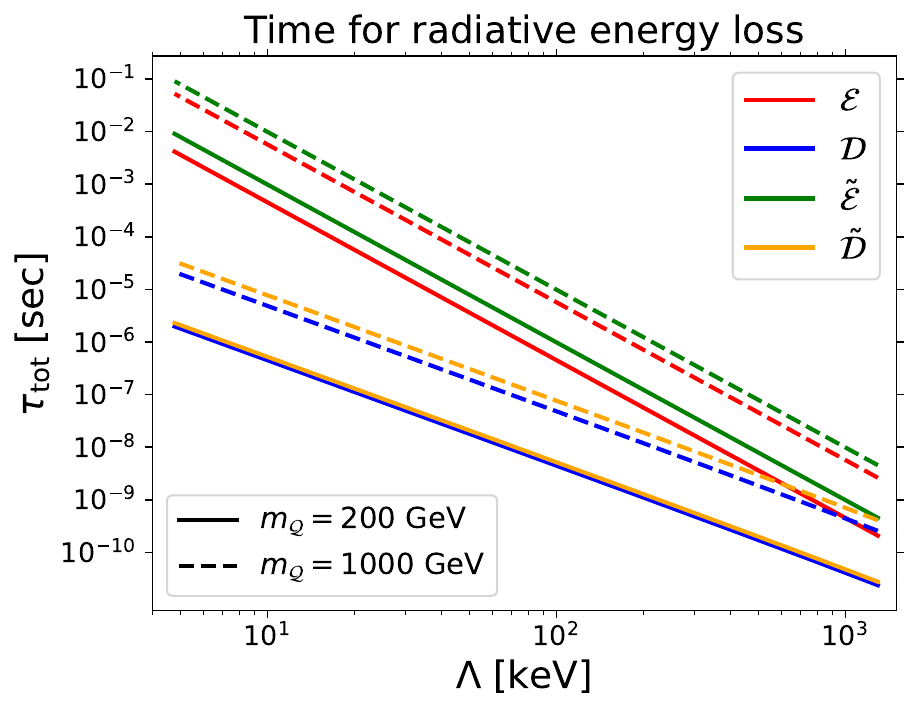}
\includegraphics[width=0.49\textwidth]{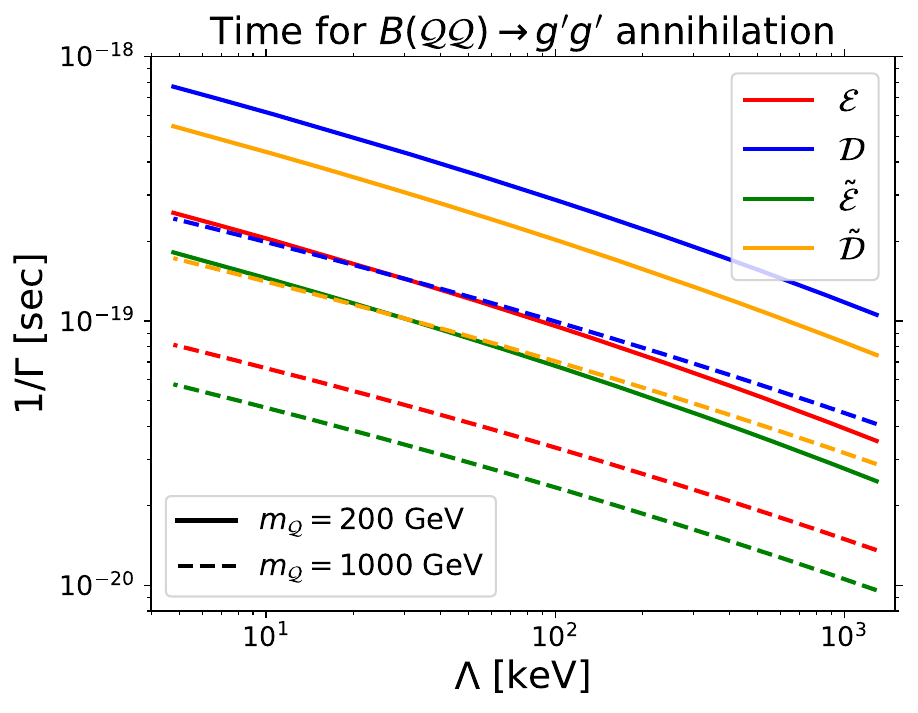}
\end{center}
\vspace*{-0.15in}
\caption{{\bf Left}:~The average time for $\mathcal{Q} \bar{\mathcal{Q}}$ states to decay to their ground state through infracolor glueball radiation with probability parameter $\epsilon = 0.1$ and, for colored quirks, QCD hadron radiation with probability parameter $\epsilon' = 0.01$. {\bf Right}:~The lifetime for $\mathcal{Q} \bar{\mathcal{Q}}$ ground states to annihilate to infracolor gluons.  For both panels, results are shown for the four quirk models with $m_{\mathcal{Q}} = 200$~GeV and 1~TeV. 
\label{fig::time}}
\end{figure}

In the left panel of Figure~\ref{fig::time}, we present the total time $\tau_{\text{tot}}=\tau^{\text{IC}}_{\text{linear}}+\tau^{\text{IC}}_{\text{Coulomb}}$ required for quirks to decay to their ground state for the four quirk models considered and two different quirk masses.  Since the decay time depends on the initial kinetic energy, which differs from event to event, we show an average time, where the average is over all events in which the quirk-pair system's momentum points into the FASER2 detector.  All quirks radiate infracolor glueballs, and the colored quirks can also lose energy through QCD hadron radiation.  Following Refs.~\cite{Evans:2018jmd, Li:2023jrt},  we choose $\epsilon=0.1$ and $\epsilon' =0.01$ for the infracolor glueball and QCD hadron radiation parameters, respectively. These parameters are not calculable from first principles; in Sec.~\ref{sec:6}, we will present results with these values varying by up to an order of magnitude.  

Despite the uncertainty in the $\epsilon$ and $\epsilon'$ parameters, we may already draw many interesting conclusions.   First, the decay time of colored quirks is far less than for color-neutral quirks: when present, QCD hadron radiation is the dominant process for shedding kinetic energy.  We also see that the time to decay to the ground state is proportional to $\sqrt{m_{\mathcal{Q}}}$. A heavier quirk will need a longer time to radiate the same amount of kinetic energy.  On the other hand, the oscillation period is reduced dramatically with increasing confinement scale.  Larger $\Lambda$ implies more rapid oscillation and, thus, a shorter time period for radiation. Last, for both masses shown, the $\tilde{\mathcal{E}}$ quirk has the longest lifetime; this is because it is color-neutral, and, at least for forward events with the $\mathcal{Q}\bar{\mathcal{Q}}$ momentum pointing toward FASER2, it also has the largest kinetic energy on average and so requires the longest time to reach its ground state. It should be noted, however, that the ordering of decay times can change if averaged over a different set of events.

{Given the lack of precise knowledge about the brown muck interaction~\cite{Kang:2008ea}, in our simulations, we have not included the deflection of the quirk-pair system by infracolor glueball and QCD hadron radiation. However, we can estimate the size of this effect. We assume isotropic and uniformly-timed radiation events, each of which carries away energy $\Lambda$ ($\Lambda_\text{QCD}$) in the quirk-pair rest frame.  A Monte Carlo simulation shows that the transverse displacement $\Delta \vec{R} = (\Delta x, \Delta y)$ of the quirk-pair in the laboratory frame at the detector due to $N_{\text{rad}}$ radiation events follows the distribution  
\begin{align}
    \text{PDF}(\Delta \vec{R}) &= \frac{1}{2\pi\sigma_R^2} e^{-\frac{\Delta x^2 + \Delta y^2}{2\sigma_R^2}} \ ,
\end{align} 
where
\begin{equation}
\frac{\sigma_R}{d_\text{IP-FASER}}  \sim  \left\{ \begin{aligned} 
 & \frac{\Lambda \times 0.34 \ N_{\text{rad}}^{1/2}}{p(\mathcal{Q} \mathcal{Q})} \sim 1.7 \ \sqrt{\epsilon} \, \left[ \frac{\Lambda}{10^7~\text{eV}}\right]^2 ~~~~\text{for }\mathcal{E} \\
 & \frac{\Lambda_{\text{QCD}} \times 0.34 \ N_{\text{rad}}^{1/2}}{p(\mathcal{Q} \mathcal{Q})} \sim 1.7 \ \sqrt{\epsilon^\prime} \, \left[ \frac{\Lambda}{10^6~\text{eV}}  \right] ~~~~\text{for }\mathcal{D} \\
\end{aligned}
\right. ~.
\label{eq:deflectionangle}
\end{equation}
In Eq.~(\ref{eq:deflectionangle}), $p(\mathcal{Q} \mathcal{Q}) \sim \mathcal{O}(1)$ TeV is the magnitude of the total momentum of the quirk-pair, $d_\text{IP-FASER} \sim 600~\text{m}$ is the distance between the ATLAS IP and the detector, and $N_\text{rad} = \epsilon N_{\text{osc}}$ $(\epsilon' N_{\text{osc}})$ for infracolor glueball (QCD hadron) radiation, where the number of oscillations is $N_{\text{osc}} \sim 25 \,[\Lambda/(100~\text{eV})]^2$ to a good approximation.

As a result, for color-neutral quirks, assuming $\epsilon \sim 0.1$, the deflection angle $\frac{\sigma_R}{d_\text{IP-FASER}} \lesssim 10^{-4}$ for $\Lambda \lesssim 10^{5}~\text{eV}$. Since the detector opening angle with respect to the ATLAS IP is at least $0.1/480$, the deflection due to radiation is negligible in the $\Lambda$ region of interest. 
On the other hand, for colored quirks, assuming $\epsilon' \sim 0.01$, the deflection can be as large as the detector opening angle for $\Lambda \lesssim 10^4~\text{eV}$. In Figure~\ref{fig::rho}, we present the quirk event density profiles in the transverse plane at the detector ($d_\text{IP-FASER} \sim 600~\text{m}$) with and without the radiation effect included. For colored quirks, large $\Lambda$ can significantly reduce the central density.  For $\Lambda=1~\text{keV}$, $10~\text{keV}$, and 50 keV, the number of events within the transverse acceptance of the FASER (FASER2) detector can be reduced by factors of 0.6, 0.15, and 0.05 (0.99, 0.69, and 0.25), respectively. 

\begin{figure}[tbp]
\begin{center}
\includegraphics[width=0.59\textwidth]{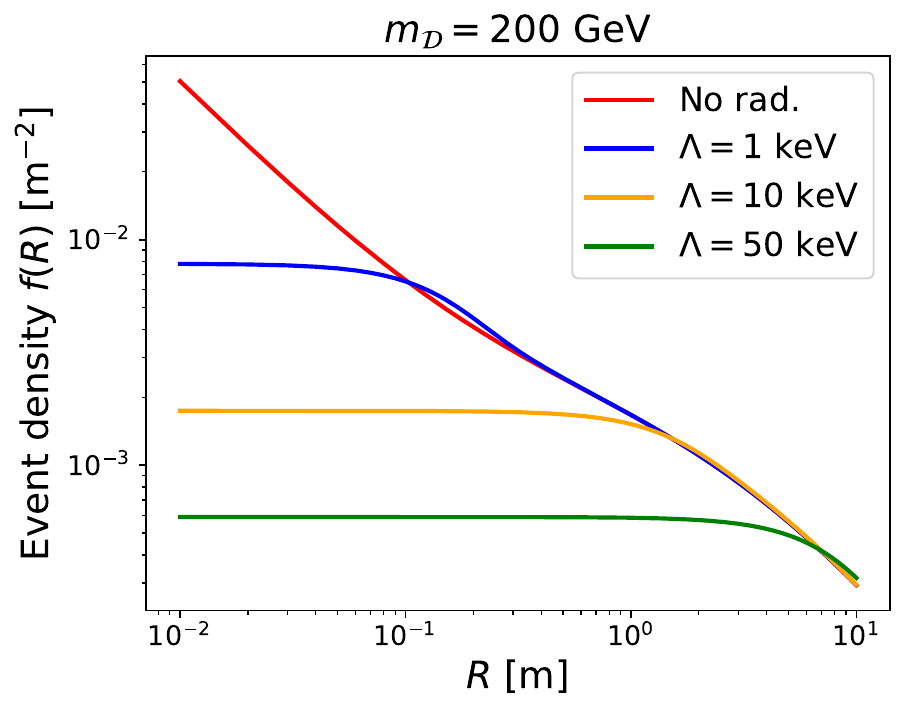}
\end{center}
\vspace*{-0.15in}
\caption{The normalized event density profile $f(R)$ for colored quirks in the transverse plane at the detector ($d_\text{IP-FASER} \sim 600$ m).  The total number of events within the FASER detector is proportional to $\int_0^{R_{\text{FASER}}} f(R) \, 2\pi R \, dR$. The red curve is the profile without the effect of radiation, while the blue, yellow, and green curves are the profiles with radiation included, with $\Lambda=1~\text{keV}$, 10 keV, and 50 keV, respectively. We have set $m_{\mathcal{D}} = 200~\text{GeV}$; the shape of the density profile has a weak dependence on the quirk mass. 
\label{fig::rho}}
\end{figure}
}

Once the quirks have shed their kinetic energy and dropped into their ground state, which we denote $B$, the ground state can decay through quirk pair annihilation.  The decay width of the ground state is~\cite{Harnik:2011mv,Barger:1987xg}
\begin{align}
\Gamma(B \to X) = \sigma v (\mathcal{Q} \bar{\mathcal{Q}} \to X) \times \left|\psi(0)\right|^2~,  
\end{align}
where $\sigma v (\mathcal{Q} \bar{\mathcal{Q}} \to X)$ is the cross section for the $S$-wave annihilation of the constituent quirks in the CoM frame, and $\psi(0)$ is the wave function of the quirkonium bound state evaluated at the origin.  In all scenarios, the quirk pair can annihilate into a pair of infracolor gluons $g'$.  The annihilation cross section for this channel in the CoM frame is
\begin{align}
\sigma v (\mathcal{Q} \bar{\mathcal{Q}} \to g^\prime g^\prime)=\begin{cases}
		\frac{\pi\alpha^2_{\text{IC}}(N_{\text{IC}}^2-1)}{4m_{\mathcal{Q}}^2 N_C N_{\text{IC}}} & \text{ for fermionic quirks} \\ 
		\frac{\pi\alpha^2_{\text{IC}}(N_{\text{IC}}^2-1)}{2m_{\mathcal{Q}}^2 N_C N_{\text{IC}}} & \text{ for scalar quirks} 
	\end{cases} ~, \label{eq:sigmaani}
\end{align}
where $N_{\text{C}} = 3$ is the number of SM QCD colors.  Other annihilation channels are also possible depending on the quirk quantum numbers. For example, colored quirks can annihilate into a pair of SM quarks or gluons, and color-neutral quirks can annihilate into pairs of SM photons or leptons. Including those channels will increase the quirkonium decay width, leading to shorter lifetimes.

The value of $\psi(0)$ is calculated by solving the stationary Schr\"{o}dinger equation with a Coulombic potential $V(r)=-\alpha_{\text{IC}}(r) / r$.  A detailed calculation can be found in Ref.~\cite{Li:2023jrt}. 
The ground state wave function depends on the quirk mass and confinement scale, because the running coupling $\alpha_{\text{IC}}$ in the quirk potential depends on these parameters. For the parameter ranges of most interest here with $m_{\mathcal{Q}} \sim 100~\text{GeV}$ and $\Lambda \sim \text{keV}$ to MeV, $|\psi(0)| \sim (10~\text{GeV})^{3/2}$ for all scenarios. 
Given the quirk annihilation cross section and the wave function at the origin, we can calculate the lifetime of the quirk ground state $\tau_{\text{ann}} = 1/\Gamma(B \to g^\prime g^\prime)$. The values for $m_{\mathcal{Q}}=200~\text{GeV}$ and 1~TeV are shown in the right panel of Figure~\ref{fig::time}. Given the quirk pair annihilation cross section of Eq.~(\ref{eq:sigmaani}), the lifetime is longer for colored fermionic quirks. Moreover, the lifetimes of quirk bound states in all scenarios exhibit a weak dependence on the confinement scale, which arises from the dependence of $|\psi (0)|$ on $\Lambda$. Note that annihilation to SM final states~\cite{Li:2023jrt} has not been included, but, of course, when possible, this will only make the annihilation time shorter.

By comparing the results for decay time and annihilation time, it can be seen that the time scale for radiative energy loss is orders of magnitude larger than that for annihilation in all scenarios.  Once the quirks decay to their ground state, the constituent quirks annihilate promptly. For quirks to be detected in FASER or future forward detectors, the quirk pair must travel at least 480 m, and so its decay lifetime must be comparable to or larger than $10^{-6}~\text{s}$.  Given the results shown in Figure~\ref{fig::time}, this requires the confinement scale to be $\Lambda \lesssim \text{MeV}$.

%%%%%%%%%%%%%%%%%%%%%%%%%%%%%%%%%%%%%%%%
\section{Quirks Traveling Toward Forward Detectors} \label{sec:5}
%%%%%%%%%%%%%%%%%%%%%%%%%%%%%%%%%%%%%%%%

From the approximate value of the quirk oscillation amplitude given in Eq.~(\ref{eq:amplitude}), we find that the amplitude is smaller than the mm scale for $\Lambda \gtrsim 3~\text{keV}$.  For this range of $\Lambda$, then, quirk events are candidate signal events if the following three criteria are met.  First, the direction of the momentum of the quirk-pair system should point into the forward detector. Second, the lifetime (which is effectively the period of radiative energy loss) of the quirk-pair system should be long enough that it can reach the detector. Third, the ionization energy loss must be small enough that the quirk pair is not stopped before it reaches the detector.  

For $\Lambda \lesssim 1$~keV, the oscillation amplitude becomes larger than the cm scale.  In this case, even if the quirk-pair momentum points into the detector, oscillations may cause one or both of the quirks to miss the detector.  On the other hand, if the quirk-pair momentum does not pass through the detector, oscillations may cause both of the quirks to pass through the detector, creating a positive signal.  For this reason, for an accurate estimate of the signal rate for $\Lambda \lesssim 1$~keV, the quirk trajectories must be modeled more precisely. 

Charged quirks travel through various materials and magnetic fields before reaching the far-forward detectors.  The quirk EoM are
\begin{align}
\frac{\partial ({m} \gamma \vec{v})}{\partial t}
&=\vec{F}_{s}+\vec{F}_{\text{ion}}~,\label{eq::move}\\
\vec{F}_{s}&=-\Lambda^2\sqrt{1-\vec{v}_{\perp}^{2}} \hat{s}-\Lambda^2 \frac{v_{ \|} \vec{v}_{\perp}}{\sqrt{1-\vec{v}_{\perp}^{2}}}~,\label{eq::fs}\\
\vec{F}_{\text{ion}}&= \frac{dE}{dx}\hat{v}~, \label{fmc}
\end{align}
where {$\gamma=1/\sqrt{1-\vec{v}^2}$,} $v_{ \|}=\vec{v}\cdot\hat{s}\nonumber$, and $\vec{v}_{\perp}=\vec{v}-v_{ \|}\hat{s}\nonumber$, with $\hat{s}$ being a unit vector along the string pointing outward at the endpoints.  $\vec{F}_s$ is the infracolor force and is derived from the Nambu-Goto action~\cite{Luscher:2002qv}.  $\vec{F}_{\text{ion}}$ is the frictional force induced by the effects of ionization energy loss.  The effects of infracolor gluon, QCD hadron, and SM photon radiation are not included in the EoM. This is justified for $\Lambda$ below the keV scale, where the quirk oscillations are less rapid. Moreover, a dedicated simulation of $R$-hadron propagation inside the detector~\cite{ATLAS:2013whh} finds that the energy lost through hadronic interactions is much smaller than the energy lost through electromagnetic ionization.  For these reasons, we do not include the energy loss due to hadronic interactions for colored quirks in the EoM.  

To simulate the quirk trajectory, various elements of the LHC infrastructure between the ATLAS IP and the forward detectors must to be considered. This includes the TAS (1.8 m long), the TAN (3.5 m long), and the concrete and rock ($\sim 100~\text{m}$ long)~\cite{Feng:2017uoz}.  The ionization energy loss of quirks inside these materials is simulated through the method described in Ref.~\cite{Li:2021tsy}.  In the HL-LHC era, the TAS and TAN will be upgraded to elements with different locations and lengths, but we have checked that this has a negligible effect on the results, given the much longer distance through concrete and rock that remains the same.   In addition, there are magnetic fields.  The quadrupole magnets tend to push the quirks toward the forward direction, leading to higher signal rates at FASER and future detectors.  The dipole magnetic field is found to change the direction of the quirk pair momentum by at most $2 \times 10^{-5}~\text{rad}$. The effects of both the quadrupole and dipole magnetic fields are highly suppressed because the quirk--anti-quirk signal is electrically neutral.  The effect of magnetic fields is therefore small, and we ignore them in our simulation for simplicity.

%%%%%%%%%%%%%%%%%%%%%%%%%%%%%%%%%%%%%%%%
\section{Timing Analyses} \label{sec:6}
%%%%%%%%%%%%%%%%%%%%%%%%%%%%%%%%%%%%%%%%

At present, the FASER detector~\cite{FASER:2022hcn} is operating on the beam collision axis at the LHC and has reported the first observations and studies of collider neutrinos~\cite{FASER:2021mtu, FASER:2023zcr, FASER:2024hoe}, as well as null results from searches for dark photons~\cite{FASER:2023tle}, U(1)$_{B-L}$ gauge bosons~\cite{FASER:2023tle}, and axion-like particles~\cite{Kock:2892328}. The SND@LHC detector~\cite{SNDLHC:2022ihg} is operating in a complementary, slightly off-axis, far-forward region at the LHC and has also observed collider neutrinos~\cite{SNDLHC:2023pun}.
Both detectors have also detected many high-energy muons and measured their fluxes~\cite{FASER:2018bac, SNDLHC:2023mib, FASER:2024hoe}.  In all of the analyses conducted so far, however, the signals of interest can be differentiated from this relatively large muon background by requiring no hits in the front veto scintillators.  

For quirks, the signal consists of charged tracks passing through the entire detector, and so alternative cuts must be devised.  As noted in Sec.~\ref{sec:intro}, we will exploit the timing information provided by the scintillator system of the FASER detector and its upgrades.  In this Section, we describe the forward detectors we will consider and provide details of the Slow Track (ST) and Delayed Track (DT) analyses, which can highly suppress the muon background, while providing satisfying signal efficiency. We then estimate the sensitivity reach of these analyses in the quirk parameter space.

\subsection{Detector Configurations}

Detailed descriptions of the FASER and FASER2 experiments, as well as the configuration of the Forward Physics Facility (FPF) experimental cavern, can be found in Refs.~\cite{FASER:2022hcn,Feng:2022inv}. 
In the following, we give a brief description of the scintillator and tracker subsystems, which are the most relevant for this study. The assumed transverse dimensions and longitudinal positions of the scintillator stations are given in Table~\ref{tab:faser}.

\begin{table}[tbp]
\centering \scalebox{1.0}{
\begin{tabular}{| c | c | c | c | c | c | c|}
\hline
   & \multicolumn{3}{|c|}{Transverse size $[\text{m} \times \text{m}]$} & \multicolumn{3}{|c|}{Longitudinal position $[\text{m}]$} \\ \cline{2-7}
 Scintillator    & FASER & FASER2  & UD & FASER & FASER2 & UD \\ \hline
Front   & 0.3~$\times$~0.3 & 1~$\times$~3 & See Figure~\ref{fig::FPF}  &  $480$ & $650$ &617 \\ \hline
Timing & 0.4~$\times$~0.4 & 1~$\times$~3& See Figure~\ref{fig::FPF} & $481.55$ & $660$  & 637 \\         \hline
Back  & 0.3~$\times$~0.3 & 1~$\times$~3 & See Figure~\ref{fig::FPF} & $484.17$ & $670$ & 682  \\            
\hline
\end{tabular}}
\caption{The transverse size and longitudinal position (distance from the ATLAS IP) of the scintillator systems of FASER, FASER2, and the UD, an ultimate detector in the FPF. \label{tab:faser}}
\end{table}

The FASER detector includes three scintillator stations, besides the veto scintillator positioned upstream of the FASER$\nu$ detector~\cite{FASER:2022hcn}. The front scintillator station is located upstream of the decay volume and is composed of two pairs of scintillators, with a 10 cm-thick lead absorber between them.  Each front scintillator counter has a transverse size of $30~\text{cm} \times 30~\text{cm}$, and is read out with a single photomultiplier tube (PMT).  Next is the timing scintillator station, which is placed right after the decay volume and in front of the tracking station.  It is made up of upper and lower scintillator counters covering the top and bottom half of the detector, respectively, with a total transverse size of $40~\text{cm} \times 40~\text{cm}$.  Each timing scintillator counter is read out with two PMTs, one on the left and one on the right.  Taking the average of the two PMT signal times allows one to reduce the uncertainty in hit time from the time-walk effect, that is, the time it takes for a signal to propagate from the hit location to a given PMT.  Finally, the back scintillator station, or pre-shower scintillator station, is placed at the back of the tracking station and in front of the calorimeter.  It is made up of two scintillator counters with a 5 cm-thick graphite absorber in between them. The transverse size of each back scintillator counter is $30~\text{cm} \times 30~\text{cm}$ and each is read out with a single PMT.  We model the FASER scintillators as centered on the LOS.  The front and back scintillators extend to angles $|\theta| \simeq 0.44~\text{mrad}$ from the beamline and pseudorapidity $\eta \simeq 8.4$.

In addition to the scintillators, the FASER detector includes 12 tracking layers, each with an active area of $24~\text{cm} \times 24~\text{cm}$, slightly smaller than the scintillators~\cite{FASER:2021ljd}. The tracking layers are composed of double-sided silicon microstrip modules with a 40~mrad stereo angle between the front and back sensors, yielding a spatial resolution of $16~\mu\text{m}$ in the vertical coordinate and $816~\mu\text{m}$ in the horizontal coordinate~\cite{FASER:2022hcn}.  We model the FASER trackers as centered on the LOS, extending to angles $|\theta| \simeq 0.35~\text{mrad}$ from the beamline and pseudorapidity $\eta \simeq 8.6$.  We will rely on the spatial resolution of the trackers to distinguish 2-track quirk events from single-track muon events.

The FASER2 detector will be located in the FPF. Its magnet aperture is currently planned to be a $1~\text{m} \times 3~\text{m}$ rectangle, and we assume that its front, timing, and back scintillators will have a similar transverse size, with their assumed distances from the ATLAS IP given in Table~\ref{tab:faser}.  As with FASER, we model the FASER2 scintillators as centered on the LOS, and they therefore extend to angles $|\theta| \simeq 2.4~\text{mrad}$ from the beamline and pseudorapidity $\eta \simeq 6.7$.  In addition, FASER2 will have multiple tracking layers.  We assume that these have the same transverse coverage as its scintillators, with a spatial resolution similar to the FASER trackers, at least in one layer of the FASER2 tracker.

\begin{figure}[tbp]
\begin{center}
\includegraphics[width=0.49\textwidth]{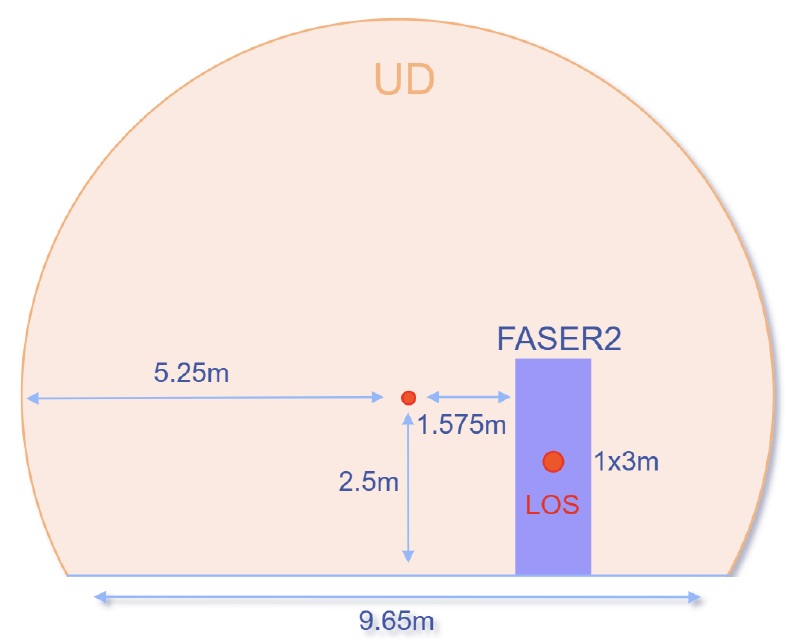}
\end{center}
 \vspace*{-0.15in}
\caption{The assumed transverse dimensions of the FASER2 and UD scintillator systems in the FPF. FASER2 has a transverse area of $3~\text{m}^2$ and extends to angles $|\theta| \simeq 2.4~\text{mrad}$ from the beamline and pseudorapidity $\eta \simeq 6.7$.  The UD has an area of $69~\text{m}^2$ and extends to angles $|\theta| \simeq 12~\text{mrad}$ from the beamline and pseudorapidity $\eta \simeq 5.1$.}
\label{fig::FPF}
\end{figure}

Last, in this study, we also consider an ``ultimate detector'' (UD) to determine the best possible sensitivity for probing quirks at the FPF. The UD is envisioned to consist of scintillators covering the entire front and back walls of the FPF cavern.  We assume that the front scintillator will also play the role of the timing scintillator. The cavern is not completely cylindrical, as there is a flat floor at the bottom, and it is not centered on the LOS. The transverse size and shape of the UD is shown in Figure~\ref{fig::FPF}, which follows the cavern configuration provided in Ref.~\cite{FASER:2022hcn}. The area of the UD scintillators is $69~\text{m}^2$, and the UD's coverage extends to angles $|\theta| \simeq 12~\text{mrad}$ from the beamline and pseudorapidity $\eta \simeq 5.1$. It is unclear how much of the UD can be instrumented with tracker.  As the purpose of the UD is to establish a theoretical upper bound on the sensitivity to quirks in the FPF, we assume that the UD has a tracker with the same transverse coverage as its scintillators and a spatial resolution similar to the FASER trackers.

\subsection{Analysis Strategy \label{sec:cuts}}

Quirks, like typical muons, appear in detectors as singly-charged MIPs.  However, quirk events have several remarkable features that can be used to distinguish them from the muon background:
\begin{itemize}
[itemsep=0.03cm, topsep=0.15cm, leftmargin=1.3em]
\item Because quirks are heavy, the typical quirk velocity is much smaller than that of the background muons. The result is hits in the scintillators that are either delayed, as determined by the arrival time at the detector, or slow, as determined by the time difference between hits in the front and back scintillators.
\item Quirk events can be observed to contain two coincident tracks, assuming the separation between the quirk and anti-quirk, characterized by the oscillation amplitude, is resolvable by the tracking detectors.
\item The quirk and anti-quirk together form an electrically-neutral, bound-state system. Their trajectories therefore do not bend in the magnetic field and appear to have very high momentum, despite the fact that they are slow.
\end{itemize}
In the following subsections, motivated by the three considerations above, we will explore the characteristics of quirk events in detail and propose cuts that preserve most of the signal but suppress the background to a negligible level. 

Note, however, that quirk events also have other distinctive features.  For example, when viewed with high resolution, quirk tracks are not helical, but the hits of the quirk tracks lie approximately in a plane. In addition, the opposite of the reconstructed momentum of the quirk-pair system points back to the ATLAS IP. In the event of a candidate quirk signal, these additional distinguishing characteristics can be used to test the quirk hypothesis.

\subsubsection{Delayed Tracks}

\begin{figure}[tbp]
\begin{center}
\includegraphics[width=0.49\textwidth]{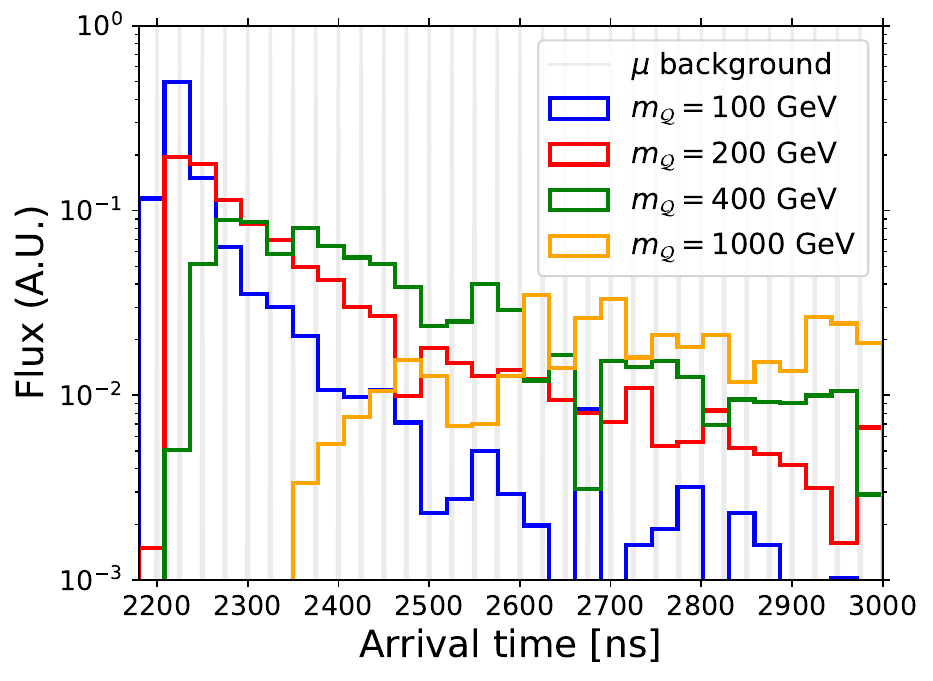}
\includegraphics[width=0.49\textwidth]{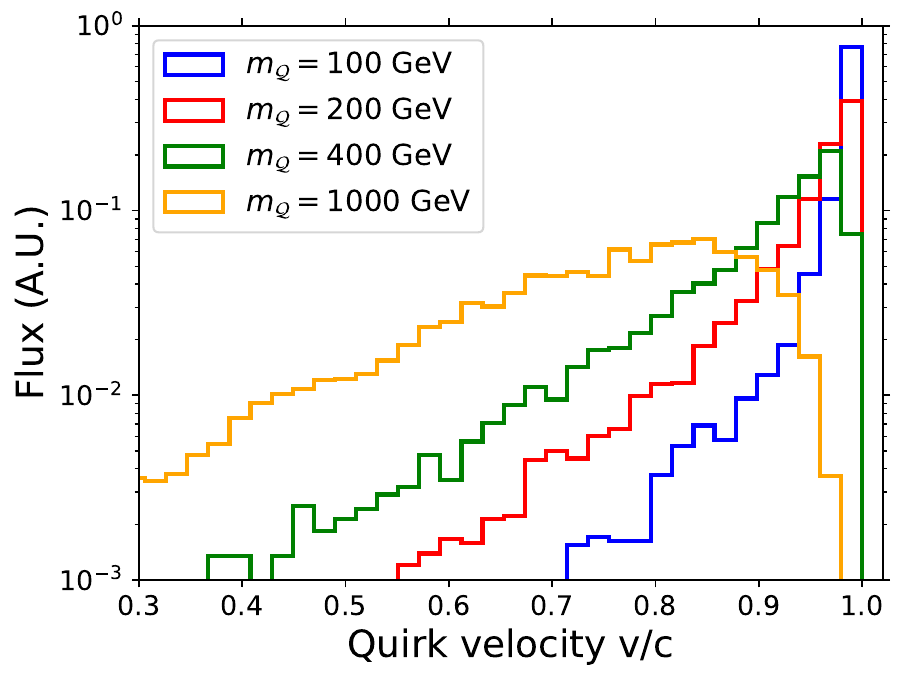}
\end{center}
\vspace*{-0.15in}
\caption{{\bf Left}:~The arrival time distributions of quirks produced at the ATLAS IP and detected in the FASER2 timing scintillator, 660~m away.  The arrival times of muons produced in bunch crossings every 25 ns are also shown, with the gray bands indicating muon timing windows of $[-3\,\text{ns}, 3\,\text{ns}]$.  {\bf Right}:~The velocities of quirks detected in the FASER2 timing scintillator.  For both panels, distributions for $m_{\mathcal{Q}} = 100$, 200, 400, and 1000 GeV are shown; the distributions are insensitive to the confinement scale $\Lambda$ and the quirk type. }
	\label{fig::quirkvelocity}
\end{figure}

At FASER, and presumably at future forward detectors, the timing scintillator has a timing resolution of approximately 500~ps, which is dominated by the time-walk effect.  This can therefore be used to measure the arrival time of particles from the ATLAS IP, and the delayed arrival of quirks at the timing scintillator can provide smoking gun evidence for the quirk signature. In the left panel of Figure~\ref{fig::quirkvelocity}, we plot the distributions of arrival times for the quirk signal and for muons, assumed to travel at the speed of light from the ATLAS IP to FASER2.  The quirk distributions strongly depend on the quirk mass, but are insensitive to the quirk type and the confinement scale. Quirks with mass larger than 100 GeV already have a significantly delayed arrival time, given the long distance from the IP to the detector, with most delayed by over one bunch crossing time (25~ns), and the effect quickly becomes more pronounced for heavier quirks. On the other hand, energetic muons simply arrive every 25~ns.  

Assuming perfect coordination with the LHC clock, the arrival time can be measured with the timing scintillator's 500~ps timing resolution.  To define a signal region, we require two coincident signal tracks to arrive more than 3~ns before or after each bunch crossing, roughly a $5\sigma$ deviation for each track.   From Figure~\ref{fig::quirkvelocity}, we see that, since the quirk arrival times are spread out over a period much larger than the bunch crossing time, such a requirement only reduces the signal by approximately $6~\text{ns} / 25~\text{ns} = 24\%$.  

The arrival time measurement mainly depends on the velocity of the quirk-pair system, while the quirk ionization energy loss is determined by the velocity of one of the quirks in the pair.  In the right panel of Figure~\ref{fig::quirkvelocity}, we plot the distribution of single quirk velocities for different quirk masses. We see that, unless $m_{\mathcal{Q}} \gtrsim 1~\text{TeV}$, very few quirks have velocity $v/c < 0.7$.  Searches for slow particles based on high $dE/dx$ are therefore much less efficient than searches based on timing.  We note also that there is a significant flux of TeV-energy muons, which lose energy through radiative processes and not just through ionization.  They are therefore not MIPs and are a background to high $dE/dx$ searches in the far-forward region. 

Given the results above, we propose the following cuts for a Delayed Track (DT) analysis, based on the arrival time of tracks in the timing scintillator:
\begin{enumerate}
[itemsep=0.03cm, topsep=0.15cm, leftmargin=1.3em]
\item A signal is detected that passes through the timing scintillator and the trackers.
\item The signal in the timing scintillator is outside the $[-3\,\text{ns}, 3\,\text{ns}]$ muon timing window.
\item The signal in the trackers is two tracks that are separated by more than $16~\mu\text{m}$ in the vertical direction.
\item The momentum of each track is greater than 100 GeV, that is, is consistent with a fairly straight track, as measured by its curvature in the magnetic field.  
\end{enumerate}

\subsubsection{Slow Tracks}

As discussed above, the FASER detector has multiple scintillators. One can therefore also look for slow tracks, that is, tracks that pass through the front and back scintillators with a time difference greater than predicted for a particle traveling at the speed of light.  Compared with the DT analysis, this ST analysis has the drawback that the distances between the front and back scintillators for FASER, FASER2, and the UD are all much shorter than the distance of these detectors from the ATLAS IP, which is in the range 480 m to 700 m. 

However, the ST analysis also has significant advantages.  First, it is independent of the LHC bunch crossing time and so requires no information from the LHC clock. Second, since quirks travel essentially parallel to the beam collision axis, the two hits in the scintillators are at the same transverse location, and so the time-walk effect cancels in the time difference.   For a MIP passing through the scintillators, preliminary data show that the time difference has a Gaussian distribution with a standard deviation of 350 ps due to random photon noise~\cite{Petersen}.  A time difference greater by 2 ns than expected for a particle traveling at the speed of light is therefore more than a $5\sigma$ discrepancy.

\begin{figure}[tbp]
\begin{center}
\includegraphics[width=0.49\textwidth]{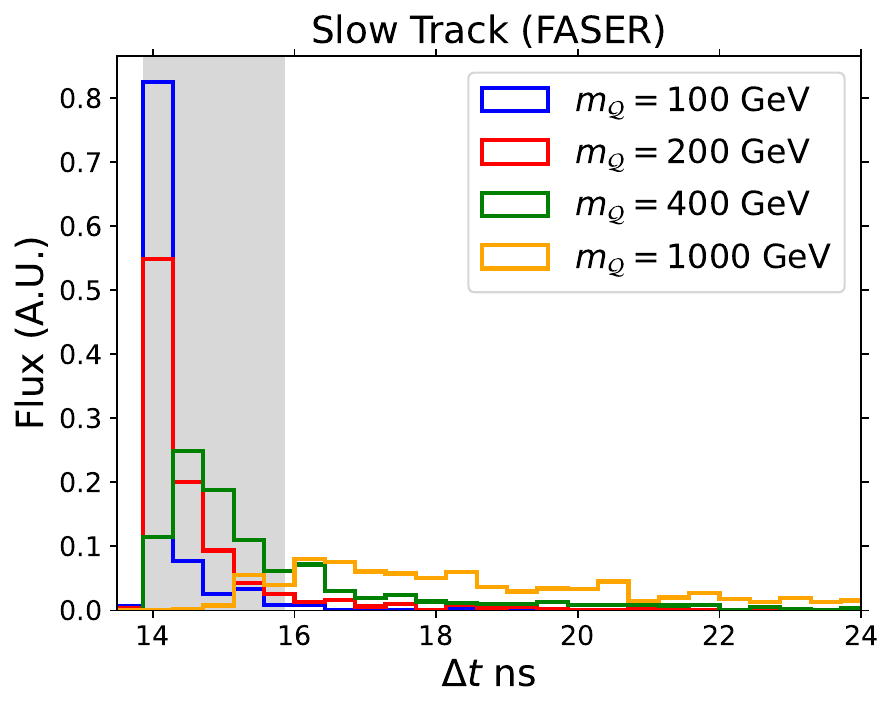}
\includegraphics[width=0.49\textwidth]{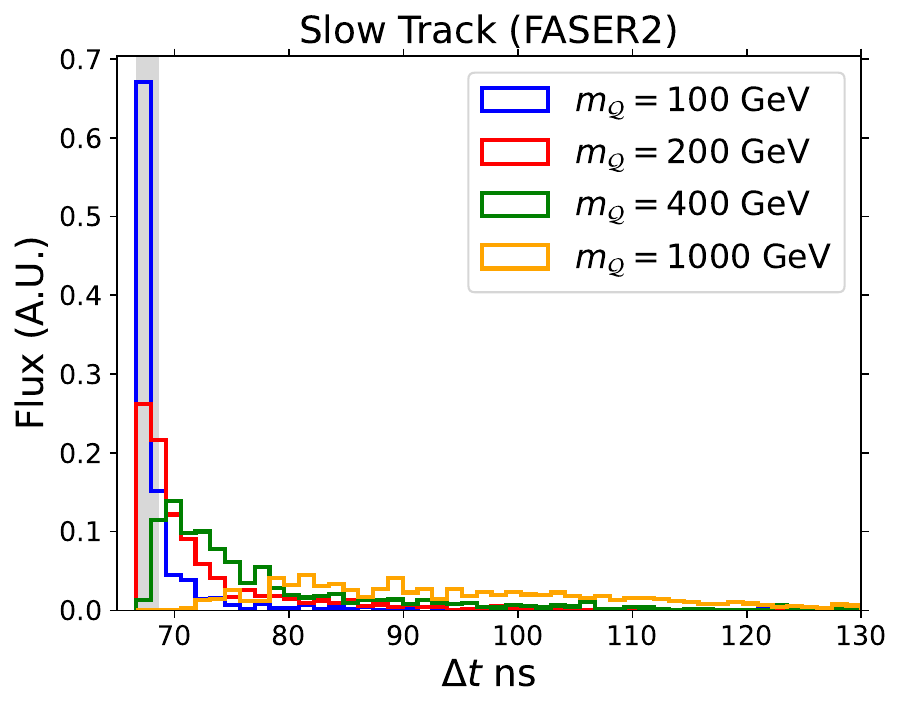}
\includegraphics[width=0.49\textwidth]{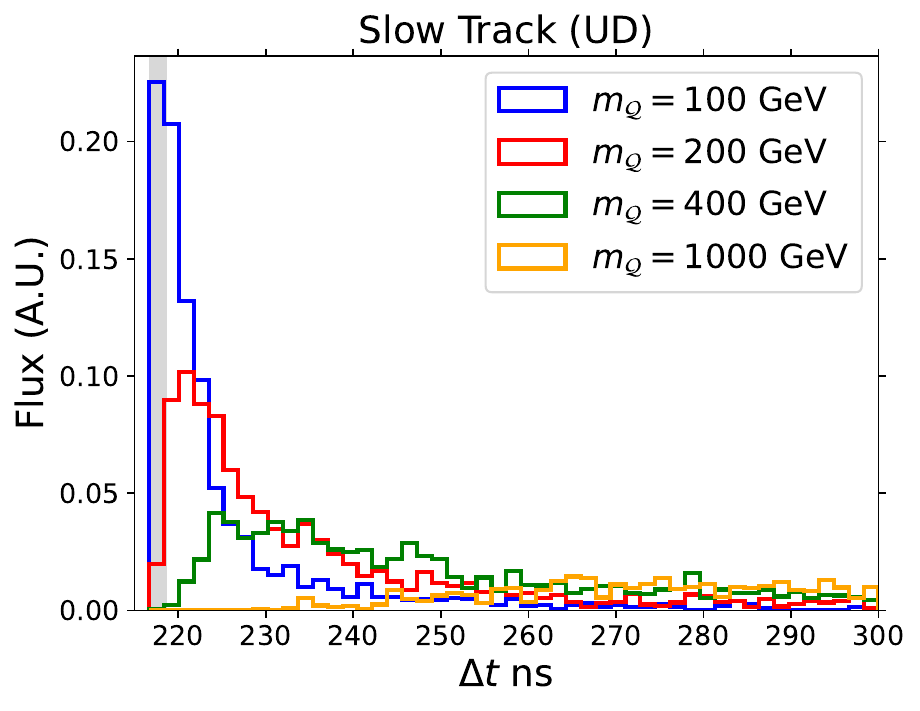}
\end{center}
 \vspace*{-0.15in}
\caption{The time difference between quirk hits in the front and back scintillators at FASER (top left), FASER2 (top right), and the UD (bottom). For all panels, distributions for $m_{\mathcal{Q}} = 100$, 200, 400, and 1000~GeV are shown; the distributions are insensitive to the confinement scale $\Lambda$ and the quirk type. Time differences outside the gray regions are more than 2~ns ($5\sigma$) greater than the time difference for muons.}
	\label{fig::timediff}
\end{figure}

The effects of such a time difference cut on the quirk signal can be seen in Figure~\ref{fig::timediff}, where we show the distribution of time differences for quirks at FASER, FASER2, and the UD. The distance between the front and back scintillators is 4.17~m for FASER, 20~m for FASER2, and 65~m for the UD, so the time difference of a muon traveling at the speed of light is 13.9~ns for FASER, 66.7~ns for FASER2, and 216.7~ns for the UD.  In the plots, the windows of time difference within 2 ns of these muons are indicated by the gray-shaded regions.   At FASER, requiring a time difference outside the 2~ns muon window will reduce the quirk signal by more than half if the quirk mass is smaller than $\sim 500~\text{GeV}$. The signal efficiencies at FASER2 and the UD are much higher due to their larger size. At FASER2, more than 50\% of events survive the selection if the quirk is heavier than 200 GeV, and almost all events pass the cut at the UD. 

Given these results, we propose another set of cuts for a Slow Track (ST) analysis based on the time difference between hits in the front and back scintillators: 
\begin{enumerate}
[itemsep=0.03cm, topsep=0.15cm, leftmargin=1.3em]
\item A signal is detected that passes through the front and back scintillators and the trackers.
\item The time difference of the hits in the front and back scintillators is more than 2~ns greater than what it would be for particles traveling at the speed of light.
\item The signal in the trackers is two tracks that are separated by more than $16~\mu\text{m}$ in the vertical direction.
\item The momentum of each track is greater than 100 GeV, as measured by its curvature in the magnetic field.   
\end{enumerate}

\subsection{Muon Background and Signal Efficiency}

A detailed experimental analysis of the backgrounds for both the DT and ST analyses is beyond the scope of this study.  However, there are good reasons to believe these analyses will be background free at both FASER and FASER2, which are sufficiently well-defined to support a preliminary evaluation of the background.

At FASER, the measured flux of tracks with energy greater than a few GeV is $(1.43 \pm 0.07) \times 10^4~\text{cm}^{-2}~\text{fb}$~\cite{FASER:2024hoe}.  Of these, almost all are muons, and FLUKA simulations imply that about half have an energy above 100 GeV~\cite{FASER:2018bac}.  The flux of muons with energy greater than 100 GeV is, then, roughly $7 \times 10^3~\text{cm}^{-2}~\text{fb}$ or, given the instantaneous luminosity of $2 \times 10^{34}~\text{cm}^{-2}~\text{s}^{-1}$, roughly $0.14~\text{Hz}~\text{cm}^{-2}$. For LHC Run 3 with an integrated luminosity of 300~fb$^{-1}$, and the tracker transverse area of $576~\text{cm}^2$, there will be a total of $1 \times 10^{9}$ muons with energy above 100 GeV passing through the FASER trackers and scintillators. 

For the ST analysis, the probability of measuring a muon with a time difference that is 1.75 ns longer than expected, a $5\sigma$ deviation, is $3 \times 10^{-7}$. For the DT analysis, we assume a similar probability for an arrival time outside the muon timing window.  The number of slow or delayed muons in all of LHC Run 3 is therefore roughly 300.  Clearly the probability of two such muons arriving at the same time is negligible.  Potentially more problematic is a single delayed or slow muon that fakes two tracks in the tracker, but the likelihood of this, especially if one requires two-track signals in multiple tracker layers, is also likely insignificant.  

For FASER2, FLUKA simulations of FPF background rates~\cite{FLUKA_FASER2} at the HL-LHC show that the expected flux of muons in FASER2 is very roughly comparable to FASER, despite the expected increase of the instantaneous luminosity to $5 \times 10^{34}~\text{cm}^{-2}~\text{s}^{-1}$. For FASER2 at the HL-LHC with an integrated luminosity of 3~ab$^{-1}$ and a tracker transverse area of $3~\text{m}^2$, then, there will be a total of $6 \times 10^{11}$ muons with energy above 100 GeV passing through FASER2.  Requiring a $5\sigma$ delay, there are roughly $2\times 10^5$ delayed or slow muons.  It again is plausible that the cuts requiring 2 tracks will be able to reduce this to a negligible background. 

Given these estimates, the dominant source of background is likely some other process, for example, delayed or slow high-energy muons that radiate a high-energy photon that pair converts into an $e^+ e^-$ pair, which then appears as two coincident tracks. Even in this case, however, for the electrons to have high enough energy to pass the cuts, they must be produced by an extremely high energy muon, which is unlikely to appear to be delayed or slow.  In summary, the signal of two coincident tracks that are delayed or slow and also high energy appears to be spectacular, and we will assume that the number of background events can be suppressed to a negligible value.

\begin{figure}[tbp]
	\begin{center}
		\includegraphics[width=0.49\textwidth]{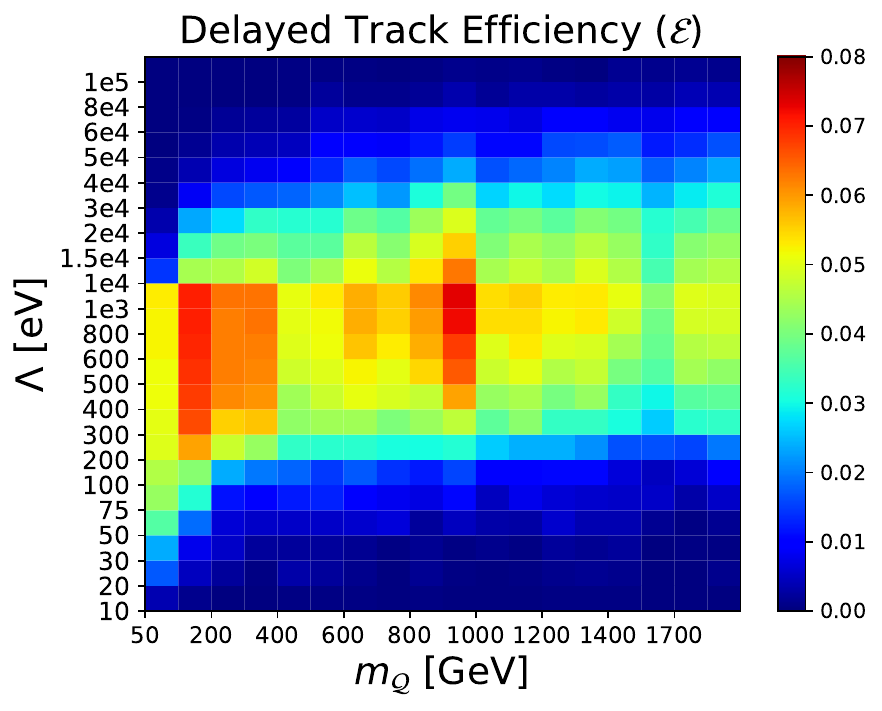}
		\includegraphics[width=0.49\textwidth]{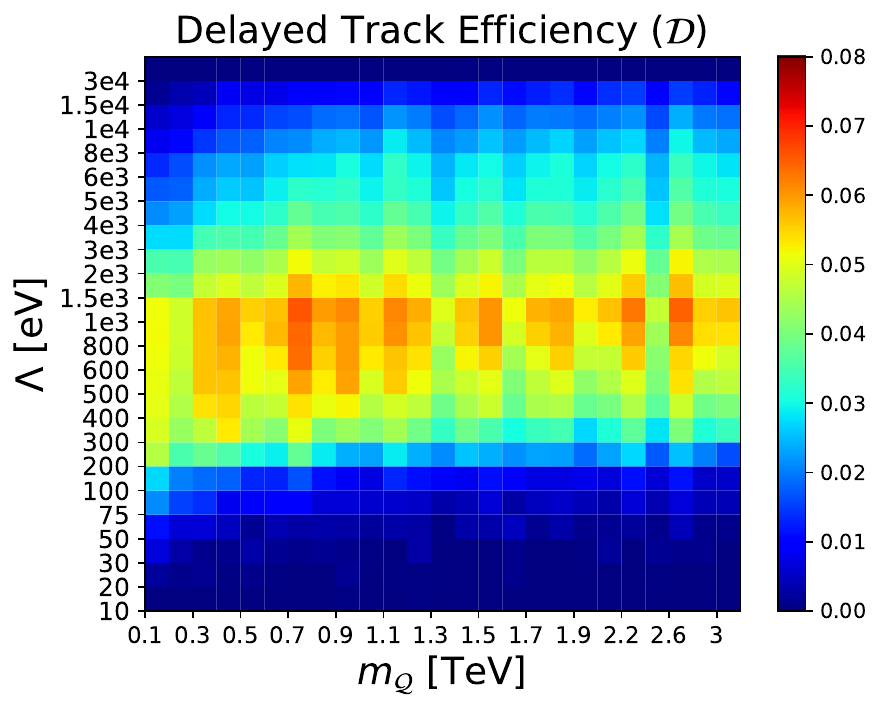}
		\includegraphics[width=0.49\textwidth]{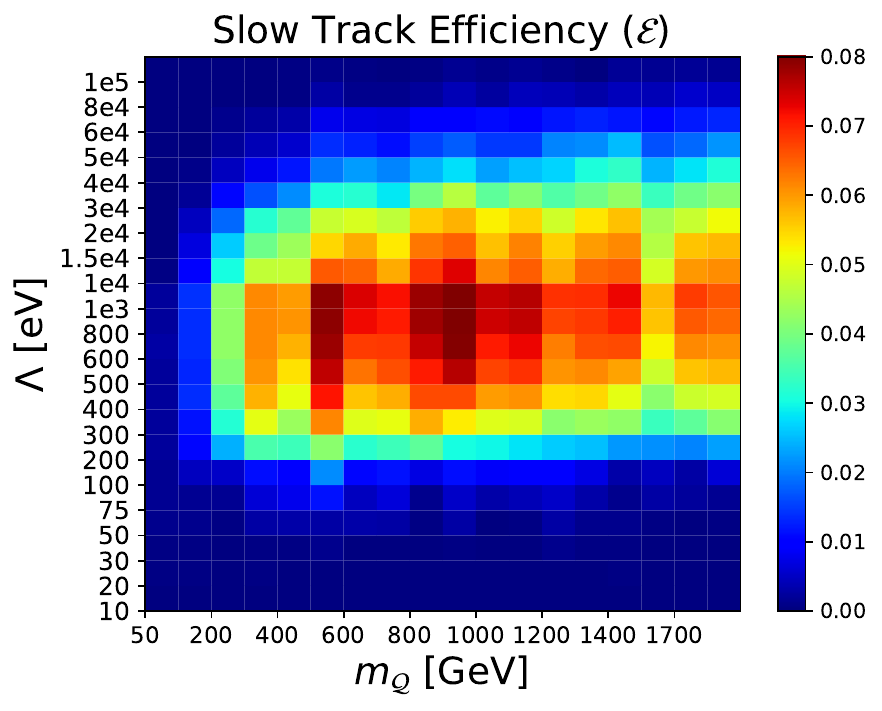}
		\includegraphics[width=0.49\textwidth]{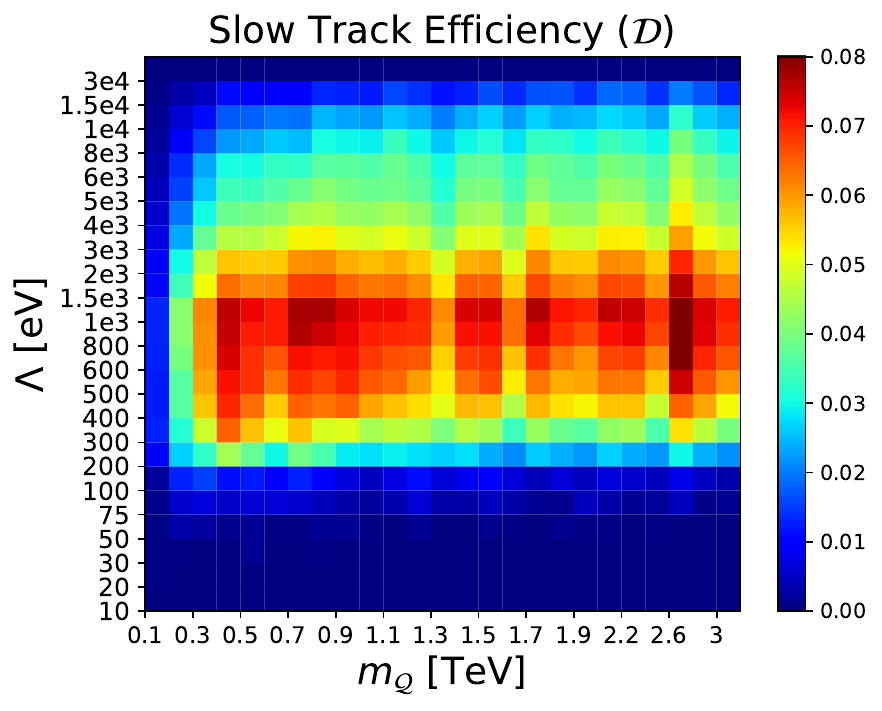}
	\end{center}
 \vspace*{-0.15in}
\caption{Efficiencies of the delayed track analysis (upper panels) and slow track analysis (lower panels) at FASER2 for the fermionic quirk models indicated. The efficiencies are the fraction of quirk events with $\thetaq < 0.005$ that pass all the cuts.  The energy loss radiation parameters are set to $\epsilon = 0.1$ and $\epsilon' = 0.01$.  Note that the $x$- and $y$-axes are not uniform; the values of $m_{\mathcal{Q}}$ and $\Lambda$ indicated represent the grid in parameter space that has been simulated. }
	\label{fig::effs}
\end{figure}

We now turn to evaluating the signal efficiency.  In Figure~\ref{fig::effs}, the signal efficiencies of the DT and ST analyses at FASER2 are shown in the $(m_{\mathcal{Q}}, \Lambda)$ plane for the two fermionic quirk models.  The main features of these distributions are similar for scalar quirks and at other forward detectors.  The fraction of quirk events with $\thetaq < 0.005$ in different scenarios can be found in the right panel of Figure~\ref{fig::effskins} with typical values around a few percent level. The efficiencies shown in Figure~\ref{fig::effs} are the fraction of quirk events that have $\thetaq < 0.005$ and also pass the cuts listed in Sec.~\ref{sec:cuts}.  The largest efficiencies found are roughly 8\%. We note that FASER2's angular coverage is up to roughly 2.4 mrad.  We have considered all events with $\thetaq < 0.005$, because quirk pairs satisfying this requirement can be detected in FASER2 when the oscillation amplitude is sufficiently large for a quirk to oscillate into the detector volume.  However, the probability for this to happen can be small for $0.002 < \thetaq < 0.005$.  Ignoring the effect of oscillations, the geometric acceptance factor can be roughly estimated to be $(3~\text{m} \times 1~\text{m})/[\pi (0.005 \times 650~\text{m})^2]$, which reduces the efficiencies plotted in Figure~\ref{fig::effs} to roughly 10\%, even before all cuts on lifetime, timing, track separation, and energy are applied.

For low $\Lambda \lesssim 100~\text{eV}$, we see that the signal efficiency becomes suppressed in both analyses. For such low $\Lambda$, the oscillation amplitude becomes large, and most events are cut by the requirement that both quirks pass through the scintillators. The only exception is for the DT analysis with $m_{\mathcal{Q}} \sim 100$ GeV, which retains a relatively high efficiency because the large oscillation amplitude allows quirks with $\thetaq>0.002$ to curve back to hit the FASER2 scintillators. 

For high $\Lambda \gtrsim 100~\text{keV}$, the efficiency is also suppressed.  For such high $\Lambda$, the oscillation amplitude is small, energy loss through radiation becomes rapid, and the requirement that the quirk pair survives long enough to reach the detector suppresses the efficiency.  

Within the range $100~\text{eV} \lesssim \Lambda \lesssim 100~\text{keV}$, the DT analysis has a much higher efficiency than the ST analysis in the low mass region $m_{\mathcal{Q}} \in [50~\text{GeV},300~\text{GeV}]$.  This is because such light quirks are not very slow, and so the ST analysis is not very efficient.  In contrast, in the DT analysis, slight velocity differences are amplified through the long distance from the ATLAS IP to the detector, and so the analysis is relatively efficient even for light quirks.  

For $\Lambda \gtrsim 1$~keV, the oscillation amplitude is much smaller than the detector size, and so only events with $\thetaq < 0.002$ can reach the FASER2 detector. Although the dependence on $m_{\mathcal{Q}}$ is clouded by fluctuations from finite Monte Carlo statistics, the fraction of forward events increases with mass for colored quirks, but decreases with mass for color-neutral quirks. As a result, for $\Lambda \sim 1$~keV, the efficiency of the DT analysis is highest in the low quirk mass region for color-neutral quirks and in the high quirk mass region for colored quirks.  However, for the ST analysis, this effect is compensated by the requirement that the quirk be heavy enough to give rise to significant time differences, as shown in Figure~\ref{fig::timediff}.

\subsection{Sensitivity Reach}

\begin{figure}[p!]
	\begin{center}
		\includegraphics[width=0.48\textwidth]{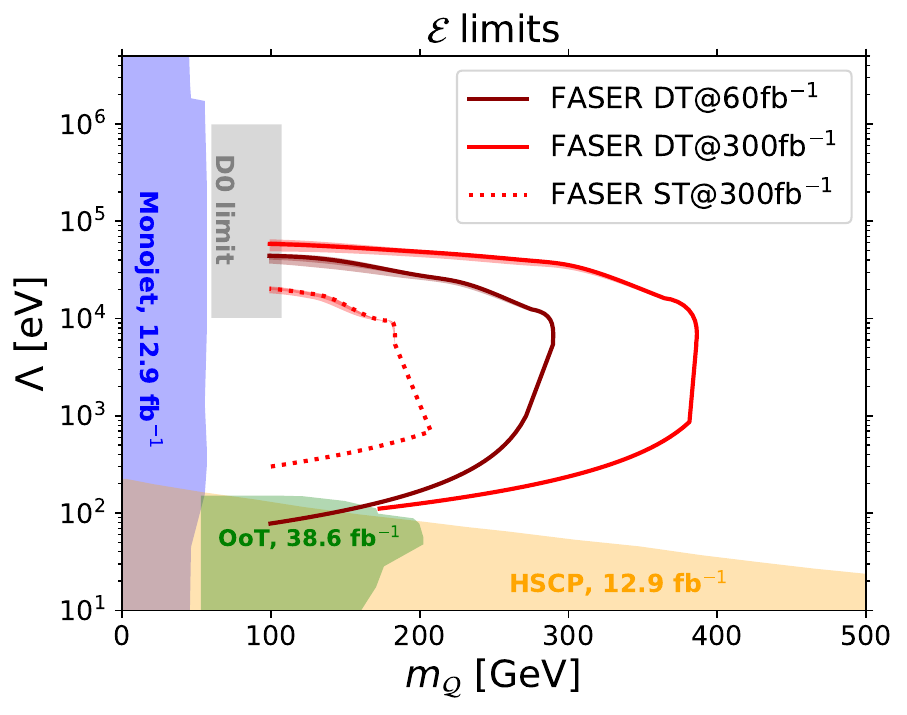}
		\includegraphics[width=0.48\textwidth]{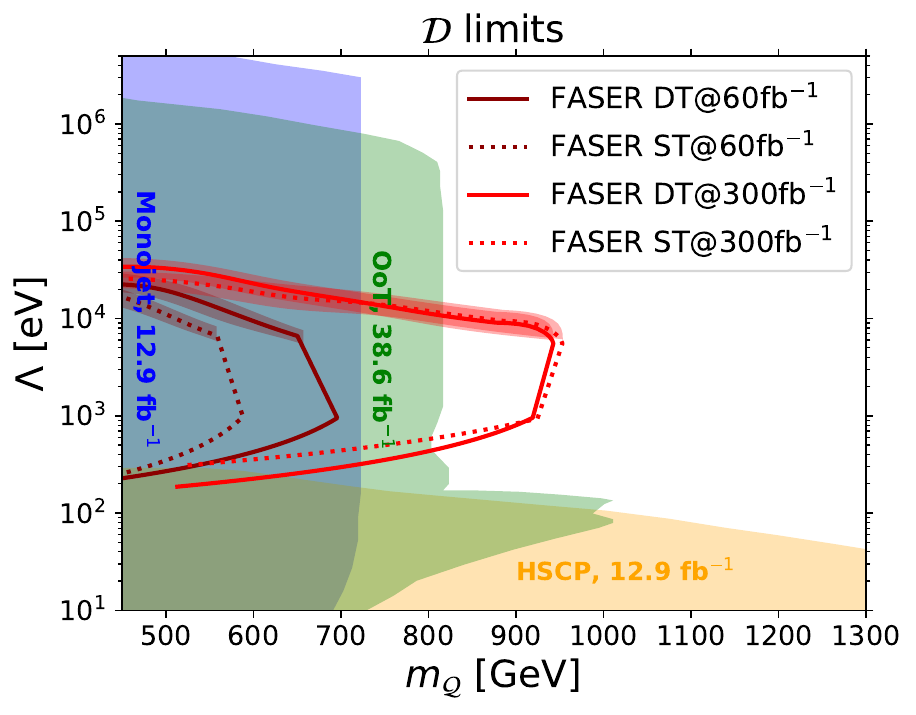}\\
		\includegraphics[width=0.48\textwidth]{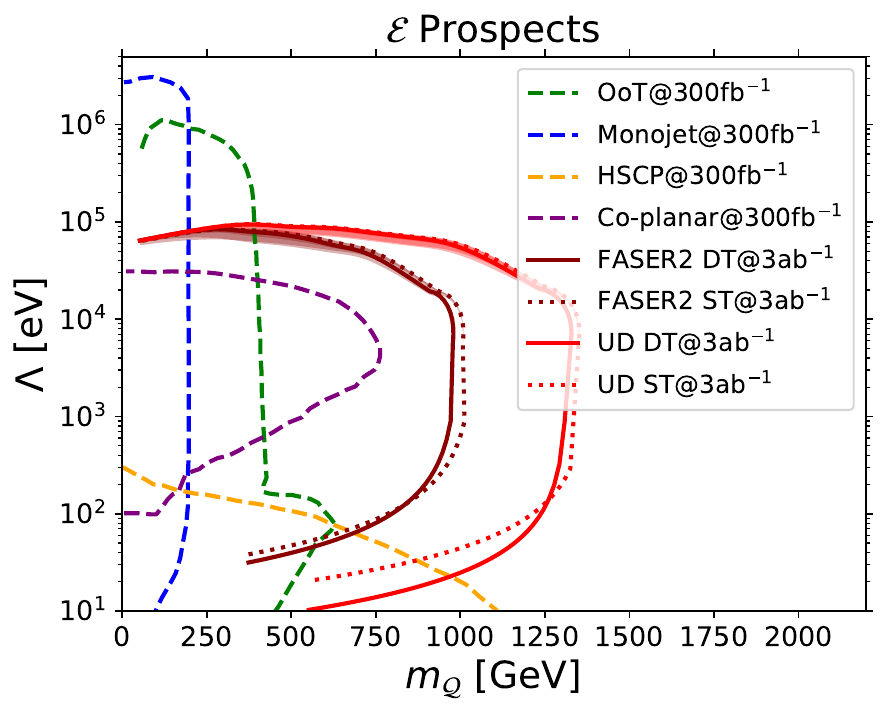}
		\includegraphics[width=0.48\textwidth]{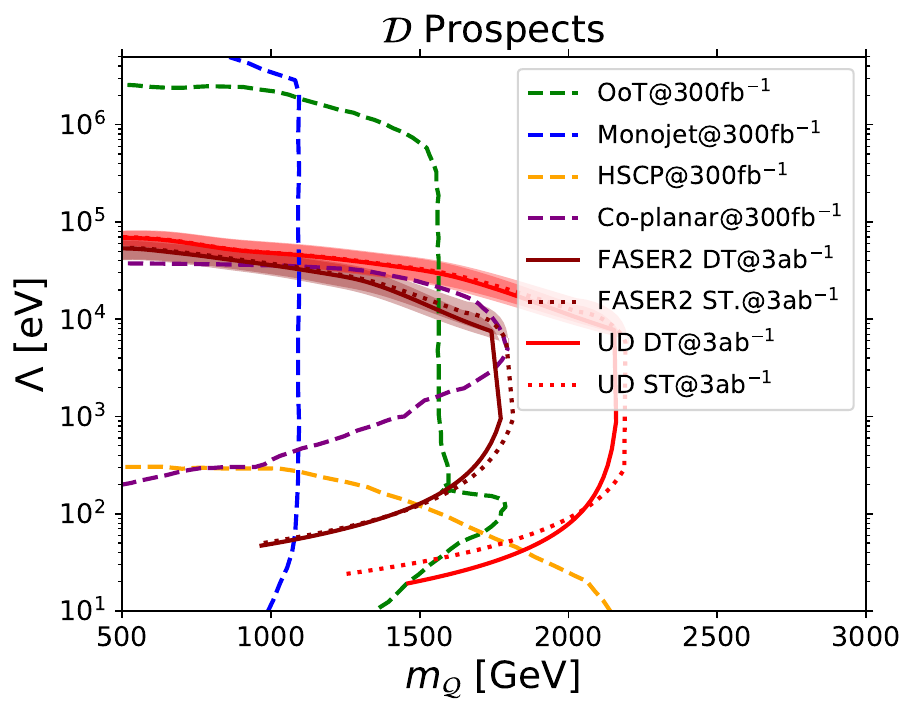} \\
		\includegraphics[width=0.48\textwidth]{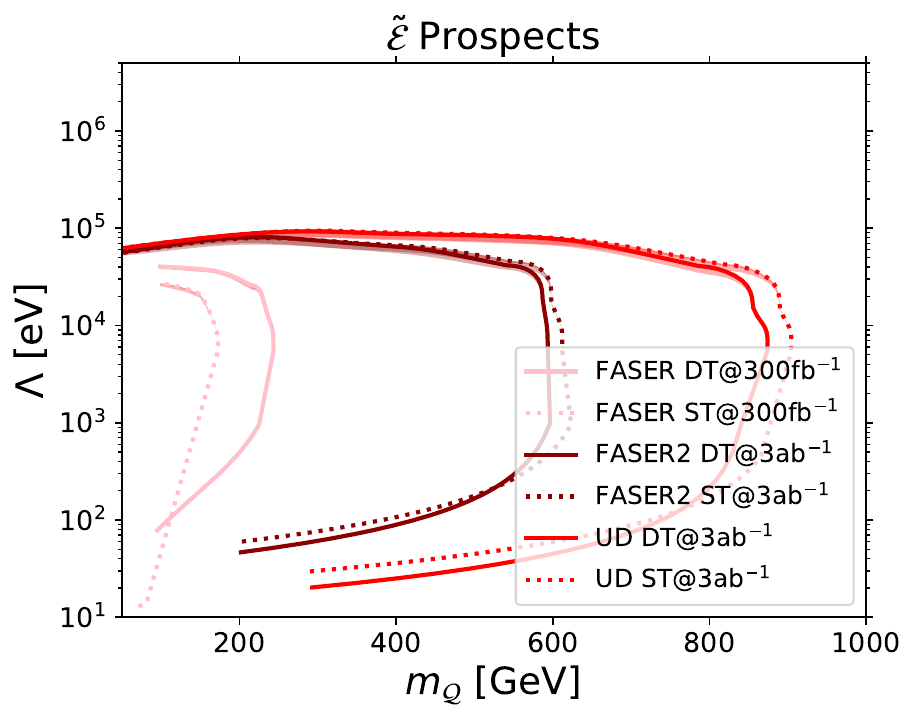}
		\includegraphics[width=0.48\textwidth]{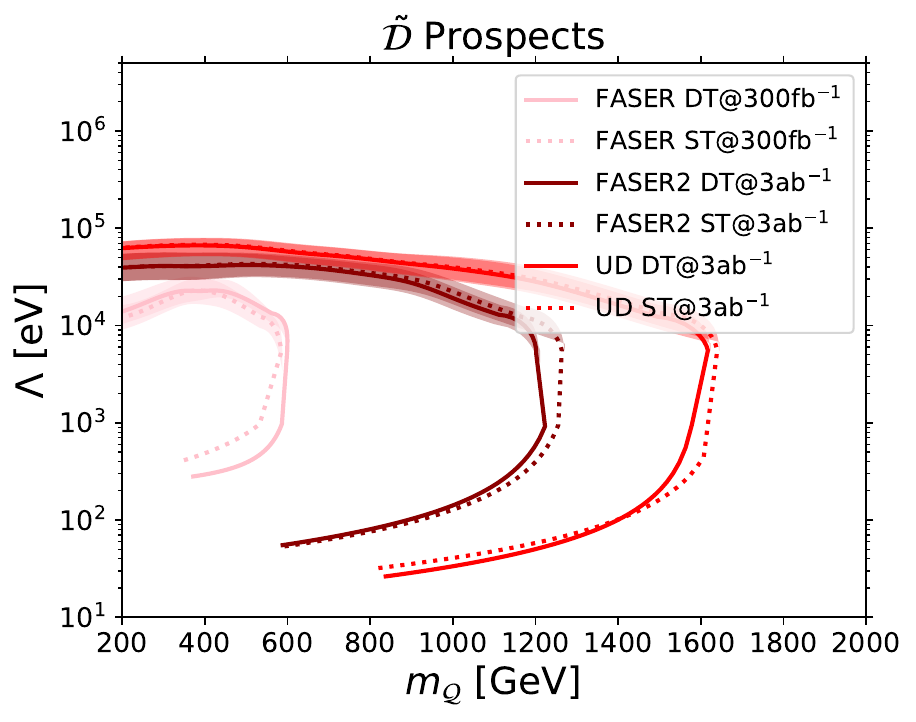} 
	\end{center}
 \vspace*{-0.15in}
\caption{{\bf Upper Panels}:~$N=3$ sensitivity contours for FASER with the current dataset ($60~\text{fb}^{-1}$) and the full Run 3 dataset ($300~\text{fb}^{-1}$) for the fermionic quirk models and the delayed track (DT) and slow track (ST) analyses.  Existing LHC bounds from OoT~\cite{Evans:2018jmd}, mono-jet~\cite{Farina:2017cts}, HSCP~\cite{Farina:2017cts}, and coplanar hits~\cite{Knapen:2017kly} searches are also shown. {\bf Middle and Lower Panels}:~$N=3$ sensitivity contours for FASER2 and the UD with the full HL-LHC dataset ($3~\text{ab}^{-1}$) for all four quirk models and the DT and ST analyses, as indicated. Also shown are projected sensitivities of searches at the LHC main detectors with $300~\text{fb}^{-1}$.  In all panels, the sensitivity contours become bands at large $\Lambda$, given uncertainties in determining the quirk-pair lifetime (see text). }
	\label{fig::exclusion}
\end{figure}

Combining the production cross sections and the selection efficiencies, we can obtain the signal rate and discovery prospects for each quirk scenario.  In Figure~\ref{fig::exclusion}, we present the $N=3$ signal event contours for the DT and ST analyses at FASER, FASER2, and the UD.  Assuming negligible background, three signal events are enough to claim a discovery.  The sensitivity reach for $\mathcal{E}$ quirks at FASER with the ST analysis and $60~\text{fb}^{-1}$ does not extend beyond a quirk mass of 100 GeV and is not shown.  For the fermionic quirks considered, the existing and projected bounds from  OoT~\cite{Evans:2018jmd}, mono-jet~\cite{Farina:2017cts}, HSCP~\cite{Farina:2017cts}, and coplanar hits~\cite{Knapen:2017kly} searches are also available in the literature, and these are shown in the relevant panels. 

In all of the cases shown in Figure~\ref{fig::exclusion}, for $\Lambda \sim 10-100$~keV, the signal rate becomes suppressed by the requirement that the quirk-pair system survives long enough to reach the detector.  The lifetime is highly sensitive to the probability of infracolor glueball ($\epsilon$) and QCD hadron ($\epsilon^\prime$) radiation. The exact values of $\epsilon$ and $\epsilon^\prime$ are determined by non-perturbative processes and are not known, but larger values of $\epsilon$ and $\epsilon'$ push the sensitivity contours to smaller values of $\Lambda$.  A more detailed discussion about the radiation probabilities and their effects on the bounds can be found in Ref.~\cite{Li:2023jrt}. To illustrate the dependence on these parameters, for color-neutral quirks, the dark-shaded (light-shaded) regions result from varying the infracolor glueball emission probability in the range of $\epsilon \in [0.03, 0.3] \ ([0.01,1])$.  For colored quirks, we fix $\epsilon=0.1$ and the shaded regions result from varying the QCD hadron radiation parameter in the range $\epsilon^\prime \in [0.005,0.02]$.

In the upper panels of Figure~\ref{fig::exclusion}, the sensitivity reaches for FASER are shown.  For the color-neutral fermionic quirk $\mathcal{E}$, even with only the currently available data, FASER with the DT analysis may probe significantly beyond current bounds for $\Lambda \in [100~\text{eV}, 40~\text{keV}]$.  With the full Run 3 data, FASER can probe new parameter space for $\Lambda \in [100~\text{eV}, 60~\text{keV}]$, extending the current bound $m_{\mathcal{Q}} > 50~\text{GeV}$ to probe quirk masses of 200 GeV with the ST analysis and 400 GeV with the DT analysis. For colored quirks, FASER will not probe as far beyond current bounds as in the color-neutral case, but FASER can still probe new parameter space with masses up to almost 1 TeV with the full Run 3 dataset in both the DT and ST analyses. 

At FASER2, the reach in quirk mass is optimal for $\Lambda \in [100~\text{eV}, 100~\text{keV}]$. 
With the full HL-LHC integrated luminosity of 3~ab$^{-1}$, FASER2 can discover quirks with masses up to 1000 GeV in the $\mathcal{E}$ scenario, 1750 GeV in the $\mathcal{D}$ scenario, 600 GeV in the $\tilde{\mathcal{E}}$ scenario, and 1200 GeV in the $\tilde{\mathcal{D}}$ scenario, thereby probing mass scales motivated by the gauge hierarchy problem and neutral naturalness models.  The ultimate reach at the FPF is given by the UD contours, which extend to quirk masses of 1300 GeV, 2200 GeV, 900 GeV, and 1600 GeV, respectively.

In the high-mass region with large $\Lambda$, the ST analysis outperforms the DT analysis because almost all of the events with the momentum direction of the quirk system pointing to the detectors are selected in the ST analysis, while around 20\% of events are vetoed by the muon windows cuts in the DT analysis, as illustrated in Figures~\ref{fig::quirkvelocity} and~\ref{fig::timediff}.   For $\Lambda$ below the keV scale, the sensitivity to large quirk masses steadily drops with decreasing $\Lambda$, which can be understood as resulting from the reduction in selection efficiency shown in Figure~\ref{fig::effs}.  In the low mass region with small $\Lambda$, the sensitivity of the ST analysis is worse than that of the DT analysis because the ST analysis requires the quirk pair to pass through both the front and back scintillators, which is more difficult to fulfill when the oscillation amplitude becomes large. 

\begin{figure}[tbp]
\begin{center}
\includegraphics[width=0.48\textwidth]{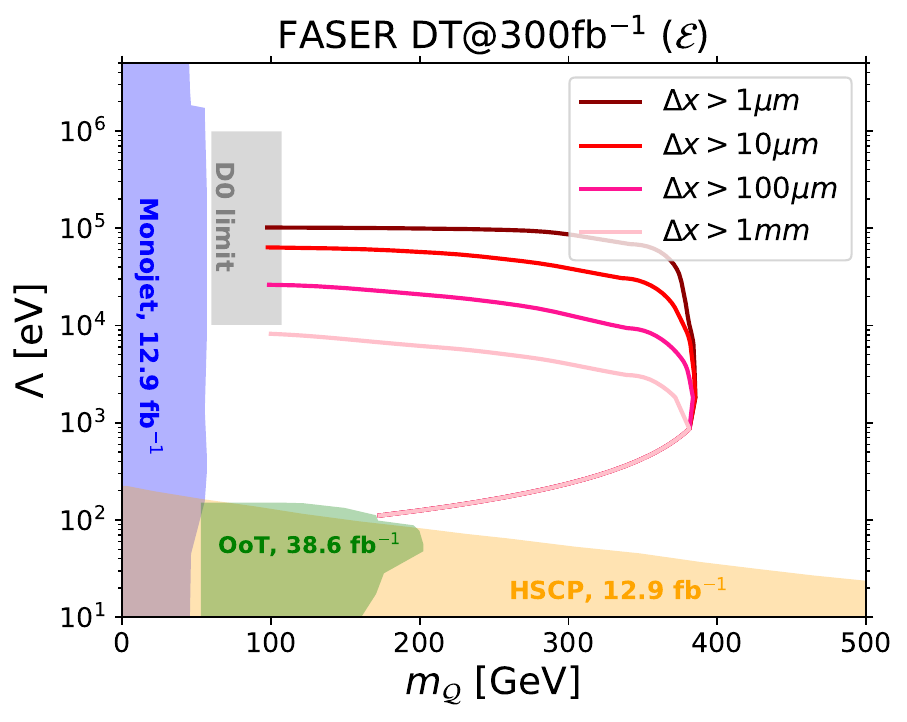} \hfill
\includegraphics[width=0.48\textwidth]{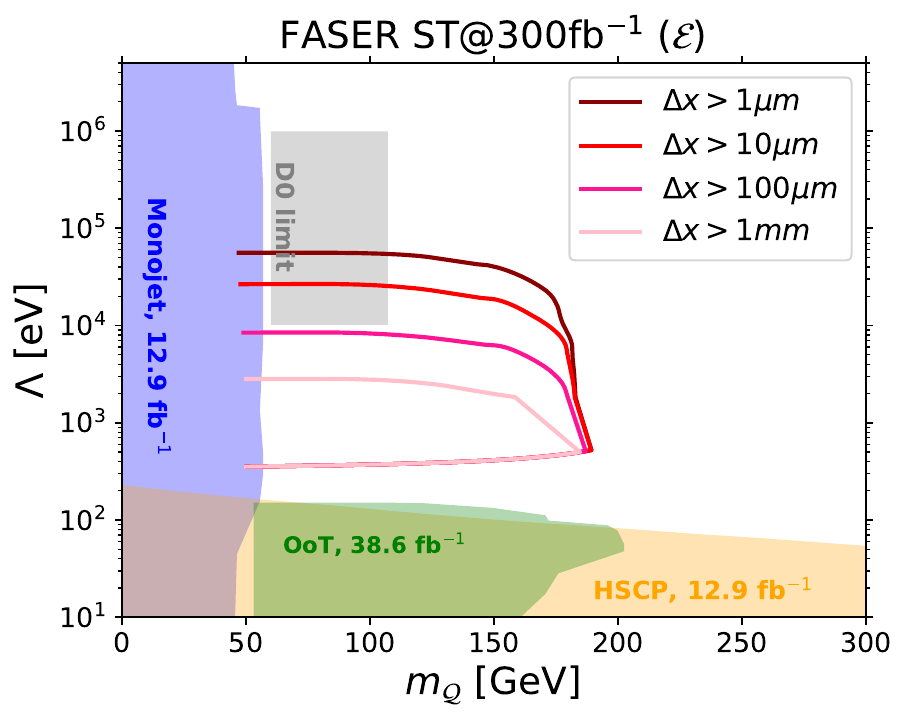}
\end{center}
 \vspace*{-0.1in}
\caption{$N=3$ sensitivity contours for FASER with the full Run 3 dataset (300 fb$^{-1}$) for the $\mathcal{E}$ quirk model and the DT (left) and ST (right) analyses.  The required separation between the quirk and anti-quirk tracks is varied from 1$\mu$m to 1 mm, as indicated. The infracolor glueball radiation probability $\epsilon$ is set to 0.1.
\label{fig::f31faser1dx}}
\end{figure}

A crucial requirement in both the DT and ST analyses is that the quirk and anti-quirk be separated by more than $16~\mu\text{m}$ so that they can be detected as two separate tracks and differentiated from the single-track muon background.  It is interesting to determine the dependence of the sensitivity on this requirement.  In Figure~\ref{fig::f31faser1dx}, we show the sensitivity reaches for the case of the $\mathcal{E}$ quirk at FASER with the full Run 3 integrated luminosity of $300~\text{fb}^{-1}$, requiring the separation to be $\Delta x = 1~\mu\text{m}$, $10~\mu\text{m}$, $100~\mu\text{m}$, and 1 mm.  For the stringent (and overly pessimistic) requirement of 1 mm separation, we see that FASER still has significant sensitivity reach beyond current bounds.  As this separation cut is relaxed, the sensitivity region grows, with the boundary at high $\Lambda$ determined roughly by $\ell \sim \Delta x$, or $\Lambda \propto (\Delta x)^{-1/2}$.  This behavior continues until $\Lambda \sim 3 \times 10^4~\text{eV}$, where the requirement that the quirk system survives long enough to travel through FASER becomes a competing constraint that reduces the efficacy of smaller $\Delta x$ in improving the reach in $\Lambda$.

%%%%%%%%%%%%%%%%%%%%%%%%%%%%%%%%%%%%%%%%
\section{Conclusions} \label{sec:7}
%%%%%%%%%%%%%%%%%%%%%%%%%%%%%%%%%%%%%%%%

This study investigates the prospects for discovering quirks at FASER, FASER2, and a possible UD using the timing information recorded by scintillators. Our simulation of quirk production at the LHC includes NLO QCD corrections. These corrections can increase the production rate by a factor of up to 2, with a more significant enhancement observed for heavier color-neutral quirks and lighter colored quirks.

For $\Lambda \lesssim 1~\text{keV}$, the simulation carefully takes into account the ionization energy loss of charged quirks traveling from the ATLAS IP to the forward detectors. For  $\Lambda \gtrsim 3~\text{keV}$, the simulation takes into account the finite lifetime of the quirk-pair system, caused by kinetic energy loss due to the radiation of infracolor glueballs and QCD hadrons during the intense quirk pair oscillations. Once the quirk-pair system settles into the quirkonium ground state, it promptly annihilates into a pair of infracolor gluons or SM particles.

We have proposed two sets of cuts to select the quirk signal: a search for two coincident delayed tracks, based on their arrival time at the timing scintillator, and a search for two coincident slow tracks, based on the time difference between their passing through the front and back scintillators.  In both cases, these cuts likely reduce the muon background to a negligible level. The DT analysis is more efficient than the ST analysis for low quirk masses $m_{\mathcal{Q}} \in [50~\text{GeV}, 300~\text{GeV}]$. 
In particular, at confinement scale $\Lambda \sim 1$~keV, the efficiency of the DT analysis is highest in a relatively low quirk mass region for color-neutral quirks and in the high quirk mass region for colored quirks. However, the ST analysis shows the highest efficiency in the high quirk mass range for all quirk scenarios.

With the $60~\text{fb}^{-1}$ of data collected in 2022 and 2023, FASER can already probe new parameter space in the case of color-neutral quirks.  With the full LHC Run 3 integrated luminosity of $300~\text{fb}^{-1}$, FASER will extend this sensitivity further into uncharted parameter space and has significant discovery prospects in both the color-neutral and colored quirk cases. 

For FASER2 at the FPF, the sensitivities of both searches significantly surpass those of projected searches at the LHC main detectors for the broad range of confinement scales $\Lambda \in [100~\text{eV}, 100~\text{keV}]$.  At FASER2 (the UD), with the full HL-LHC integrated luminosity of 3~ab$^{-1}$, we can discover quirks with masses up to 1000 (1300) GeV in the $\mathcal{E}$ scenario, 1750 (2200) GeV in the $\mathcal{D}$ scenario, 600 (900) GeV in the $\tilde{\mathcal{E}}$ scenario, and 1200 (1600) GeV in the $\tilde{\mathcal{D}}$ scenario, probing parameter space motivated by the gauge hierarchy problem.

\begin{acknowledgments}
We thank Aki Ariga, Tomoko Ariga, Jamie Boyd, Hide Otono, and especially Brian Petersen for providing many insights and essential guidance regarding experimental aspects of this work.  This work was supported by the Natural Science Foundation of Sichuan Province under grant No.~2023NSFSC1329 and the National Natural Science Foundation of China under grant Nos.~11905149 and 12247119.  The work of JLF was supported in part by U.S.~National Science Foundation Grant Nos.~PHY-2111427 and PHY-2210283, Simons Investigator Award \#376204, Heising-Simons Foundation Grant Nos.~2019-1179 and 2020-1840, and Simons Foundation Grant No.~623683. 
\end{acknowledgments}

\bibliographystyle{jhep}
\bibliography{quirk_faser}
\end{document}